\documentclass{aa}

\usepackage{amsfonts}       
\usepackage{graphicx}
\usepackage{natbib}
\usepackage{textcomp}
\usepackage{xcolor}
\usepackage{hyperref}

\newcommand{\pd}[2]{\frac{\partial #1}{\partial #2}}

\newcommand{\Mm}{~\mathrm{Mm}}
\newcommand{\cm}{~\mathrm{cm}}
\newcommand{\erg}{~\mathrm{erg}}

\newcommand{\km}{~\mathrm{km}}

\newcommand{\ev}{~\mathrm{ev}}

\title{Chromospheric Extension of the MURaM Code}
\author{ D. Przybylski\inst{1}, R. Cameron\inst{1}, S.K. Solanki\inst{1,2}, M. Rempel \inst{3}, J. Leenaarts \inst{4}, L. S. Anusha\inst{1}, V. Witzke\inst{1}, A.I. Shapiro\inst{1}}
\authorrunning{Przybylski et al.}
\begin{document}
\institute{Max-Planck Institute for Solar System Research, 37077 G\"{o}ttingen, Germany
            \mail{przybylski@mps.mpg.de}
            \and
            School of Space Research, Kyung Hee University, Yongin, Gyeonggi 446-701, Republic of Korea
            \and
            High Altitude Observatory, NCAR, P.O. Box 3000, Boulder, Colorado 80307, USA
             \and
            Institute for Solar Physics, Dept. of Astronomy, Stockholm University, AlbaNova University Centre, SE-10691 Stockholm, Sweden
          }

\abstract{Detailed numerical models of chromosphere and corona are required to understand the heating of the solar atmosphere. An accurate treatment of the solar chromosphere is complicated by the effects arising from Non Local Thermodynamic Equilibrium (NLTE) radiative transfer. A small number of strong, highly scattering lines 
dominate the cooling and heating in the chromosphere. Additionally, the recombination times of ionised hydrogen are longer than the dynamical timescales, requiring a non-equilibrium (NE) treatment of hydrogen ionisation.}
{We describe a set of necessary additions to the MURaM code so that it might handle some of the important NLTE effects. We investigate the impact on models of the solar chromosphere caused by NLTE and NE effects in radiation magnetohydrodynamic \textbf{(rMHD)} simulations of the solar atmosphere.}
{The MURaM code is extended to include the physical process required for accurate simulation of the solar chromosphere, as implemented in the Bifrost code. This includes a time-dependent treatment of hydrogen ionisation, a scattering multi-group radiation transfer scheme and approximations for NLTE radiative cooling.}
{The inclusion of NE and NLTE physics has a large impact on the structure of the chromosphere; the NE treatment of hydrogen ionisation leads to a higher ionisation fraction and enhanced populations in the first excited state throughout cold inter-shock regions of the chromosphere. Additionally this prevents hydrogen ioniation from buffering energy fluctuations, leading to hotter shocks and cooler inter-shock regions. The hydrogen populations in the ground and first excited state are enhanced by $10^2-10^3$ in the upper chromosphere and up to $10^9$ near the transition region.}
{Including the necessary NLTE physics leads to significant differences in chromospheric structure and dynamics. The thermodynamics and hydrogen populations calculated using the extended version of the MURaM code are consistent with previous non-equilibrium simulations. The electron number and temperature calculated using the non-equilibrium treatment of the chromosphere are required to accurately synthesise chromospheric spectral lines.}

\maketitle

\section{Introduction}
The importance of the solar chromosphere for resolving a number of the large open questions in solar physics is undoubted. E.g., it provides the connection between the solar surface, the source of the energy for the upper solar atmosphere, and the corona, where this energy is deposited and the local plasma heated, as well as the solar wind accelerated. A detailed understanding of the dynamics and structure of the solar chromosphere has consequently been a major goal of solar physics for decades. Energy transfer in the chromosphere is strongly affected by the interplay between radiation and the chromospheric plasma. An accurate treatment of radiation transfer is necessary to model the structure of the chromosphere.

The theoretical treatment of radiation transfer \textbf{(RT)} in the  chromosphere is difficult because it cannot be treated as optically thin as in the corona, nor can it be treated in local-thermodynamic equilibrium (LTE), as in the photosphere. Additionally, the large recombination timescale of ionised hydrogen and helium mean the problem cannot be treated in statistical equilibrium (SE).  The dynamics and radiation effects must be solved together in non-equilibrium \citep{Carlsson_2002_dynamichydrogen,judge_2005_NE_atomic_populations}. Additionally, the low ionisation fraction and low collisional frequencies may lead to a  drift between the ionised and neutral component of the plasma. The weak coupling between ions and neutrals can lead to ambipolar diffusion and Hall drift becoming significant, or even require a multi-fluid treatment.

A multi-dimensional treatment of the chromosphere is complicated by the non-locality of NLTE radiation transport. The important chromospheric spectral lines are strongly scattering and should in principle be treated with partial frequency redistribution (PRD). For NLTE RT simulations, the computational time scales proportionally to $n_z^2$, where $n_z$ is the number of points in the vertical direction. The 3D spectral synthesis of a single chromospheric spectral line, such as Ca H\&K, Mg h\&k, or hydrogen Lyman-Alpha can cost 50-200 kcore-H per million grid-points\citep{sukharov_2017_multi3D_PRD}. The computational cost of detailed NLTE radiation transfer ($23.23~\mathrm{s}$ per grid point for hydrogen in PRD) compared to an update of the multi-group LTE rMHD code ($15.80~\mu \mathrm{s}$ per grid point), leaves large multi-dimensional simulations including detailed radiative NLTE transfer out of reach. One dimensional codes exist, for example the RADYN code \citep{carlsson_1992_RMHD}. In multi-dimensional rMHD simulations of the solar chromosphere, two NLTE effects are critical. The radiative cooling due to NLTE chromospheric spectral lines, and the rate of ionisation/recombination of hydrogen and helium in the solar chromosphere. For multi-dimensional simulations, we turn to approximations of these effects in order to make 3D simulations computationally tractable.

The 3D computational modelling of the solar choromosphere has been so far led by the Bifrost group \citep{Gudiksen_2011_Bifrost}. The Bifrost code includes a number of approximations to NLTE and NE physics, producing the most realistic 3D simulations of the chromosphere currently available. The approximations include tabulated recipes for computationally efficient NLTE chromospheric line losses, based on detailed synthesis of the radiation field including PRD effects \citep{carlsson_2012_approximations}, scattering multi-group radiation transfer \citep{skartlien_2000_multigroup,hayek_2010_radiative}, and a NE hydrogen Equation of State (EoS) \citep{leenaarts_2007_nonequilibrium}. To more accurately simulate the transition region, Bifrost includes a 3D treatment of hydrogen Lyman lines and the addition of a computationally efficient helium model atom \citep{golding_2016_NEhelium}.

In this paper we introduce an updated version of the MURaM code \citep{voegler_2005_muramcode,rempel_2014_numerical,rempel_2017_extension}  which includes prescriptions for NLTE and NE effects. The prescriptions used are the same as those described in the Bifrost code \citep{Gudiksen_2011_Bifrost}, but do not include the extensions of \citet{golding_2016_NEhelium}.

The MURaM code has been employed to treat many phenomena in the solar photosphere, such as umbral dots \citep{schussler_2006_umraldots}, sunspots \citep{Rempel_2009_sunspot}, small-scale dynamo \citep{Voegler_2007_dynamo}, magnetic flux emergence \citep{Cheung_2007_fluxemergence,Chen_2017_fluxemergence}, etc.  and has been used to investigate the effect of ambipolar-diffusion in 3D rMHD simulation \citep{Cheung_2012_muram_PI,danilovic_2017_3DEBs}. The code has been used to compute molecular \citep{Schuessler_2003_MURaMmolecular}, and atomic \citep{Shelyag_2007_MURaMstokesdiagnostics} diagnostics, as well as for modelling of the solar irradiance variability \citep{Shapiro_2017_brightnessvariation,Yeo_2017_irradiance}.  It has also successfully treated phenomena in the solar corona \citep{rempel_2017_extension,Chueng_2019_flare} and in stellar photospheres \citep{Beeck_2013_stellarconvection,beeck_2015_fluxconcentrations,Panja_2020_starspots}. With the extension described here, this versatile code will close one of the large remaining gaps in its ability to treat phenomena in the solar atmosphere, namely covering the solar chromosphere.

 In Sect. \ref{sec:numerics} we describe the numerical methods used, including the diffusion scheme, NE equation of state and radiative losses. In Sect. \ref{sec:setup} we outline the experimental setup and present the first simulations of the solar chromosphere using this updated version of MURaM. In Sect. \ref{sec:results_rad} we analyze the radiative cooling in the chromosphere, and in Sect. \ref{sec:results_pops} we discuss the effect of the NE treatment of hydrogen in the chromosphere. Finally in Sect. \ref{sec:discussion_conclusion} we discuss the results and present our conclusions.

\section{Numerical Approach}\label{sec:numerics}

The MURaM code \citep{voegler_2005_muramcode} solves the conservative MHD equations on a Cartesian grid in one-, two- or three-dimensions. Spatial derivatives are calculated using a fourth-order central difference scheme. Temporal integration is performed with the Jameson-Schmidt-Turkel scheme \citep{jameson_2017_timeupdate}, a 4-stage explicit time-update scheme. The code's original hyperdiffusion scheme was replaced by a hybrid scheme based around slope-limiters and higher order hyperdiffusion scheme \citep{rempel_2014_numerical}. Further enhancements were made by \citet{rempel_2017_extension} to allow simulations of the solar corona, including;  optically thin losses, point-implicit heat conduction and a semi-relativistic `Boris correction' to circumvent the time step restrictions due to the high Alfv\'en velocity \citep{Boris_1970_BC}. A pre-tabulated EoS using the Opal \citep{rogers_1996_OPAL}, or Uppsala \citep{Gudiksen_2011_Bifrost} packages is used to calculate temperature, pressure and electron number from the plasma density and internal energy. 

 In this section we describe further extensions to realistically simulate the solar chromosphere, including modifications to the diffusion scheme, the Equation of State and the implementation of a non-equilibrium treatment of hydrogen populations, and radiative cooling and heating.

We first summarise the procedure used to update the system of equations over a time-step $\Delta t$. First the radiative heating and cooling (Sect \ref{sec:radiation}) are calculated using the system state at the previous time $t_0$. In each sub-stage $n$ the steps followed are: calculate the right-hand-side (RHS) of the MHD equations using a directionally unsplit approach (Sect. \ref{sec:MHD}), integrate in time to the new state $t_* = t_0 + 1/(5-n) \Delta t$, then apply the boundary conditions, advect the populations, and evaluate the EoS (Sect. \ref{subsec:EoS}). No hyperdiffusion or explicit diffusivities are included in the code, instead once all the sub-stages have been completed the directionally split diffusion scheme (Sect. \ref{subsec:diffusion}) is run. After each directional sweep the EoS equations must be solved. Next the hyperbolic $\nabla \cdot B$ cleaner \citep{Dedner_2002_divB} is applied. After the MHD variables have been updated to the next timestep, the EoS is solved once more, including a set of hydrogen rate equations (Sect \ref{sec:NE_hydrogen}).

\subsection{Numerical Diffusion Scheme}\label{subsec:diffusion}

The MURaM code includes a hybrid diffusion scheme which is based around slope-limiters and includes higher order hyper-diffusion terms \citep{rempel_2014_numerical,rempel_2017_extension}. For the simulations presented in this work, we introduce an additional scheme based around the Partial Donor Cell Method (PDM) as described in detail in \citet{zhang_2019_gamera}. The PDM limiter is used to avoid undershoot or overshoot that occurs near discontinuities when a high-order (HO) scheme is used.

To apply the limiter to a quantity $\Phi$, the flux through the interface $i+1/2$ between cell $i$ and $i+1$ must be calculated. We use a 4th order \textbf{centred} reconstruction to calculate the high-order value at the cell interface $\Phi_{i+1/2}^{\mathrm{HO}}$. When a discontinuity is detected, the PDM limiter decides whether the value at the cell interfaces needs to be "limited". The left ($l$) and right ($r$) interface values are calculated.

\begin{eqnarray}
\Phi^l_{i+1/2} &=& \Phi^{HO}_{i+1/2} - \mathrm{sign}\left(\nabla^+ \right) \mathrm{max} \left[0,|\Phi^{HO}_{i+1/2}-\Phi_i| \right.\label{eqn:pdm_left}\nonumber\\
&-& \left. C_{\mathrm{pdm}} \left(\nabla^{-}\nabla^{+} 
- |\nabla^{-}\nabla^{+}|\right)|\nabla^{-}| \right],~\mathrm{and} \\
\Phi^r_{i+1/2} &=& \Phi^{HO}_{i+1/2} + \mathrm{sign}\left(\nabla^+ \right) \mathrm{max} \left[0,|\Phi_{i+1}-\Phi^{HO}_{i+1/2}| \right.\label{eqn:pdm_right}\nonumber\\
&-& \left. C_{\mathrm{pdm}} \left(\nabla^{++}\nabla^{+}
- |\nabla^{++}\nabla^{+}|\right)|\nabla^{++}| \right],
\end{eqnarray}
where $\nabla^+ = \Phi_{i+1}-\Phi_i$, $\nabla^- = \Phi_{i}-\Phi_{i-1}$, $\nabla^{++} = \Phi_{i+2}-\Phi_{i+1}$. $C_{\mathrm{pdm}}$ is a parameter that controls the amount of diffusion, with $C_{pdm}=0$ equivalent to a first order donor cell scheme, and $C_{pdm} > 0$ corresponding to a lower diffusivity. The diffusive flux $f_{i+1/2}$ across the cell interface is calculated from the limited values
\begin{eqnarray}
\label{eqn:diff_flux}
f_{i+1/2} &=& -\frac{c_{i+1/2}}{2} |\mathrm{sign}\left(\Phi^r_{i+1/2}-\Phi^l_{i+1/2}\right) \nonumber\\
&+& \mathrm{sign}\left(\nabla^+\right)|\left(\Phi^r_{i+1/2}-\Phi^l_{i+1/2}\right),
\end{eqnarray}
where  $c_{i+1/2}= c_{\mathrm{s}} + v_{\mathrm{A}} + |\mathbf{v}|$ the characteristic velocity in terms of the sound speed $c_{\mathrm{s}}$, Alfv\'en velocity $v_{\mathrm{A}}$ and velocity vector $\mathbf{v}$. Once the fluxes are calculated, the remainder of the scheme is the same as that described in \citet{rempel_2017_extension}. The diffusion scheme is applied to the logarithm of the density $\rho$, internal energy per gram $\epsilon$, and population fractions $\mathbf{n}/\rho$, as well as the velocity components $v_x,v_y,v_z$, and magnetic field vector $B_x,B_y,B_z$. Applying the diffusion to the logarithm in a stratified atmosphere reduces the systematic vertical diffusive fluxes. The energy, momentum and population numbers are corrected for mass diffusion.

To increase stability and minimise diffusion, a number of enhancements were made, following those in \citet{rempel_2017_extension}. In order to remove wiggles superposed on the stratified atmosphere fourth order hyper diffusion is added in the vertical direction to $\log \rho, \log \epsilon,\log \mathbf{n} ~\mathrm{and}~v_z$.

 Additionally, a number of switches exist to allow the code to run at a less diffusive setting, while taking care of the few, localised gridpoints which require higher diffusion. One can operate the scheme without these switches, but a more diffusive setting would be required everywhere in order to keep the code stable. The default value of diffusion coefficient in the simulation is $C_{\mathrm{pdm}}=2$. These switches include; To reduce the errors in $\nabla \cdot B$ the diffusive flux of $B_z$ in the z-direction is set to zero at the vertical boundaries and the numerical diffusivity of $\mathbf{B}$ in the direction of $\mathbf{B}$ is reduced by a factor of 0.2. Secondly, to reduce the diffusion in the convection zone the sound-speed contribution to the characteristic velocity $c_{i+1/2}$ is limited to a maximum of $1~\mathrm{km\;s}^{-1}$.

Two hard switches are included to prevent the formation of instabilities which can occur in the solar atmosphere. Firstly, if the maximum density contrast between grid-points exceeds 10, then the mass diffusivity is increased ($C_{\mathrm{pdm}}=0$). This setting ensures stability of the code, providing extra diffusion at sharp shock-fronts and other extreme events. Secondly, diffusivity in regions with a low adiabatic index ($\gamma<1.12$) is increased ($C_{\mathrm{pdm}}=0.5$). The increased diffusivity at low gamma was included due to an instability in the chromosphere. When a strong flow is present, and the plasma is around the ionisation temperature of hydrogen (the dominant species), recombination of protons to neutral hydrogen can cause a sharp drop in pressure. The pressure gradient, combined with the diffusion scheme, was found to drive an instability at the grid-scale. The diffusion scheme used is not a smooth Laplacian, the numerical diffusivities are highly intermittent. The addition of these two extra switches does not greatly change the behaviour of the diffusion scheme, but allows for the use of an overall lower diffusivity in the numerical domain. 

The MURaM code, like many solar and stellar rMHD codes, use non-isotropic grid spacing in the horizontal and vertical direction, many codes (e.g. Bifrost \citep{Gudiksen_2011_Bifrost}) additionally use a non-uniform spacing in the vertical direction. The ratio of horizontal to vertical grid spacing is typically around $1.5-3$. This ratio will lead to anisotropies in the numerical diffusion. In simulations of the convection zone and atmosphere, the structures modelled are highly anisotropic. Even the use of an isotropic numerical grid will not give isotropic diffusivities. A higher vertical resolution is suitable for convection simulations, as the vertical-to-horizontal ratio of convective cells is approximately $1/3$. Although this argument breaks down in the atmosphere, the increased vertical resolution is also advantageous for radiative transport. The grid anisotropy allows better resolution of sharp vertical gradients, such as those present at the photosphere and transition region. Simulations with the MURaM code and a variety of different resolutions have found no significant effects due to the grid anisotropy \citep{Voegler_2007_dynamo,rempel_2014_numerical}. Comparison of these simulations and others with non-uniform meshes \citep{Beeck_2012_muram_co5bold_stagger_compare} have found no significant systematic differences with using the anisotropic grid. An additional systematic effect of the diffusivity can exist due to a tendency for vertical diffusive fluxes in the stratified atmosphere. The systematic effects in the horizontally averaged diffusive fluxes of mass, energy and vertical momentum remain small at the photosphere. We find that the horizontally averaged hydrostatic balance is conserved to within a percent in the convection zone and lower atmosphere. In the corona, large cross-field gradients of thermodynamic quantities can exist, which can lead to enhanced diffusive fluxes.

\subsection{Non-Equilibrium Equation of State}\label{subsec:EoS}

To perform self-consistent simulations of the solar atmosphere we require an equation of state (EoS). We combine two approaches; firstly, we use a pre-tabulated LTE equation of state to model the interior based on the prescription introduced by \citet{vardya_1965_thermodynamics}, and extended by \citet{Mihalas_1967_calculation} and \citet{Wittmann_1974_computation} (VMW), see also \citet{Vitas_2015_VMW}. Secondly, an equation of state including a non-equilibrium treatment of hydrogen is used for the solar chromosphere and corona. The pre-tabulated LTE EoS is used for pressure greater than $2\times10^5 \;\mathrm{dyn\;cm}^{-2}$, and an EoS state including the non-equilibrium treatment of hydrogen is used for lower pressures.  As the non-ideal gas and non-equilibrium ionisation effects are negligible at this pressure the two methods join smoothly. We use a mixture of the 15 most abundant elements of the Sun and include the molecules $H_2$ in non-equilibrium, and $H_2^+$ and $H^{-}$ in chemical equilibrium. We include up to three ionisation states for all LTE elements and a 5-level plus continuum model of hydrogen in NE.  In this Section we focus on the NE EoS, an overview of the LTE EoS is given in Appendix \ref{sec:LTE_EOS}.

At each sub-stage of the iteration scheme, or after each directional-sweep of the diffusion scheme, the MHD solver provides updated values of the internal energy density $E_{\mathrm{int}}$ and density $\rho$ of the plasma. From the density and the atomic abundances, the total number density of hydrogen nuclei $n_{\mathrm{H,tot}}$ and non-hydrogen nuclei $n_{\mathrm{nonH,tot}}$ are calculated. We then find a solution to the equations of energy conservation, charge conservation and nuclei conservation, in terms of temperature ($T$), electron number density ($n_{\mathrm{e}}$) and population number densities ($\mathbf{n} = \left\{n_{a,i,j}\right\}$). We use the notation $(a,i,j)$ to represent a species (atom or molecule) $a$ of ionisation stage $i$ and energy level $j$.

Two tables are required for the NE EoS to include the contribution from non-hydrogen atoms. The thermodynamics of non-hydrogen atoms are treated in LTE. The electron number density per hydrogen nuclei ($n_{\mathrm{e,nonH}}$) and energies of excitation and ionisation per hydrogen nuclei ($\epsilon_{\mathrm{nonH}}$) for non-hydrogen atoms are tabulated as a function of temperature and electron number. These are calculated as;
\begin{equation}
n_{\mathrm{\mathrm{e,nonH}}}\left(n_{\mathrm{e}},T\right) = \sum_{a=1}^{15} \sum_{i=0}^{3} i\frac{n_{a,i}}{n_{a,\mathrm{tot}}} \frac{n_{a,\mathrm{tot}}}{n_{\mathrm{H,tot}}},
\end{equation}
\begin{equation}
\epsilon_{\mathrm{nonH}}\left(n_{\mathrm{e}},T\right) = \sum_{a=1}^{15} \sum_{i=0}^{3} \left[\chi_{a,i} + k_{\mathrm{B}} T^2 \pd{\ln\left(U_{a,i}\right)}{T}\right]\frac{n_{a,i}}{n_{a,\mathrm{tot}}}  \frac{n_{a,\mathrm{tot}}}{n_{\mathrm{H,tot}}},
\end{equation}

Where $n_{a,\mathrm{tot}}/n_{H,\mathrm{tot}}$ is the number fraction of element $a$ relative to hydrogen, $\chi_{a,i}$ is the ionisation energy, $k_{\mathrm{B}}$ the Boltzmann constant, $U$ is the partition function  and $n_{a,i}/n_{a,\mathrm{tot}}$ is the fraction of element $a$ in ionisation stage $i$ calculated using Saha-Boltzmann (Eqn. \ref{eqn:sahaboltzmann}). The partition functions used are described in Appendix  \ref{sec:LTE_EOS}.

Following \citet{leenaarts_2007_nonequilibrium} the NE EoS is evaluated by solving a system of equations using a Newton-Raphson method. The equation of energy conservation is
\begin{eqnarray}
f_0 &=& 1 - \frac{1}{E_{\mathrm{int}}} \left(\frac{3 k_{\mathrm{B}} T}{2}\left[n_{\mathrm{e}} + n_{\mathrm{nonH}} + n_{\mathrm{H}_2} + n_{\mathrm{H}_2,1} + n_{\mathrm{H},-1} \right.\right.\nonumber\\
&+& \left.\left. \sum_{i,j} n_{_{\mathrm{H},i,j}}\right] + n_{\mathrm{H,tot}} \epsilon_{\mathrm{nonH}} + n_{\mathrm{H}_2} E_{\mathrm{H}_2} + n_{\mathrm{H}_2,1} E_{\mathrm{H}_2,1} \right. \label{eqn:NE_energy}  \nonumber\\
&+& \left. n_{\mathrm{H},-1} E_{\mathrm{H},-1} + \sum_{i,j} n_{_{\mathrm{H},i,j}} E_{_{\mathrm{H},i,j}} \right) = 0,
\end{eqnarray}
where the $E_{a,i,j}$ gives the ionisation, excitation and dissociation energies of the atom/molecule. The derivatives of $n_{\mathrm{\mathrm{e,nonH}}}$ and $\epsilon_{\mathrm{nonH}}$ with respect to $n_{\mathrm{e}}$ and $T$ are calculated numerically from the table interpolants. The energies of the hydrogen species are given by Eqns. \ref{eqn:E_H2}-\ref{eqn:E_H}. The equation of charge conservation is
\begin{equation}
\label{eqn:NE_charge}
f_1 = 1 - \frac{1}{n_{\mathrm{e}}} \left(n_{\mathrm{H},1}+n_{\mathrm{H}_2,1}-n_{\mathrm{H},-1} + n_{\mathrm{H,tot}} n_{\mathrm{e,nonH}}\right) = 0.
\end{equation}
Finally, nucleus conservation must be maintained. All non-hydrogen elements are considered in LTE.
The equation of hydrogen nucleus conservation is 
\begin{equation}
\label{eqn:NE_Hcons}
f_2 = 1 - \frac{1}{n_{\mathrm{H,tot}}} \left(\sum_{i,j} n_{_{\mathrm{H},i,j}} + 2 n_{\mathrm{H}_2} + 2 n_{\mathrm{H}_2,1} + n_{\mathrm{H},-1} \right) = 0.
\end{equation}
A full set of the derivatives are described in Appendix \ref{sec:NE_EOS_ders}. 

\subsection{Non-equilibrium hydrogen populations}\label{sec:NE_hydrogen}

The time evolution of a species $n_{a,i,j}$ depends on the advection of the populations with the bulk fluid velocity $\mathbf{v}$, the rate of collisional ($C_{ij,kl}$) and radiative ($R_{ij,kl}$) transitions between level $\left(ij\right)$ and level $\left(kl\neq ij\right)$, and the rate of molecule formation or destruction by gas-phase reactions $(r)$. The gas-phase reactions are described in terms of a set of reactants ($A,B,C$) and rate coefficient $K_r$ which form ($r^+$) or form ($r^-$) the species $\left(aij\right)$. Defining $P_{ij,kl} = C_{ij,kl} + R_{ij,kl}$, the rate equation of a species $n_{aij}$ is described by
\begin{eqnarray}
\label{eqn:population_evolution}
\pd{n_{aij}}{t} &=& - \nabla \cdot \left(n _{aij}\mathbf{v}\right) \nonumber\\
&+&  \sum_{kl \neq ij} n_{ak} P_{kl,ij} - n_{aij} \sum_{kl \neq ij} P_{ij,kl} \nonumber \\
&+&  \sum_{r^+} n_{r^+,A} n_{r^+,B}n_{r^+,C} K_{r^+} \nonumber\\
&-& \sum_{r^-} n_{r^-,A} n_{r^-,B}n_{r^-,C} K_{r^-}.
\end{eqnarray}
In this work we treat only hydrogen and $H_2$ in non-equilibrium. To solve Eqn. \ref{eqn:population_evolution} we solve separately for the continuity and the rates. The advection of the species by the macroscopic fluid velocity $\mathbf{v}$ is then given by
\begin{equation}
\pd{n_{aij}}{t} = - \nabla \cdot \left(n_{aij}\mathbf{v}\right).
\end{equation}
We found that advecting the populations with the 4th order central differences and 4-stage temporal integration scheme used for MHD variables leads to frequent negative values. Instead, we use a flux-limited unsplit donor cell method. Each sub-stage of the temporal integration scheme advances the fluid variables $\rho,\epsilon,\mathbf{v}$ in time, with a timestep $\Delta t$, from $t_0$ to $t_{*} = t_0 + \Delta t$. To advect the populations we use the average velocity  $\mathbf{v}^{1/2} = 0.5\left(\mathbf{v}^{t_0} + \mathbf{v}^{t_{*}}\right)$. For each direction, the velocities $\mathbf{v}^{1/2}$ are interpolated to the cell interfaces using quadratic Bezier interpolation \citep{delacruz_2013_DELOBezier}. The limited left(l)- and right(r)-interface values are calculated using the PDM limiter (Eqn. \ref{eqn:pdm_left} \& \ref{eqn:pdm_right}), as applied in the diffusion scheme (Sect. \ref{subsec:diffusion}). The flux of the quantity $\Phi$ through the cell interface $i+1/2$, in direction $d=x,y,z$, is calculated as

\begin{eqnarray}
f_{d,i+1/2} &=& \frac{1}{2} (v^{1/2}_{d,i+1/2}+|v^{1/2}_{d,i+1/2}|) \Phi^l_{i+1/2} \nonumber\\
&+&\frac{1}{2} (v_{d,i+1/2}^{1/2}-|v_{d,i+1/2}^{1/2}|) \Phi^r_{i+1/2},
\end{eqnarray}
and the populations are updated using the sum of the fluxes through all faces of the cell,
\begin{equation}
\Phi^{t_{*}} = \Phi^{t_0} + \sum_{d=x,y,z} \frac{\Delta t}{\Delta d} \left(f_{d,i-1/2}-f_{d,i+1/2}\right) .
\end{equation}
Once the advected populations $n^*_{a,i,j}$ have been calculated, an update of the EoS Eqns. \ref{eqn:NE_energy}-\ref{eqn:NE_Hcons} ensures consistency between the EoS variables, populations, and magnetohydrodynamic energy $E_{\mathrm{int}}^*$ and density $\rho^*$. After all sub-stages are complete the directionally split diffusion scheme is run. The populations are diffused using the scheme described in Section $\ref{subsec:diffusion}$. The populations are also corrected for any mass diffusion.

In order to evaluate the system of hydrogen rates concurrently with the EoS, the rate equations are written in a form suitable for solution with the Newton Raphson method,
\begin{eqnarray}
f_{3+i,j} &=& \frac{n_{aij}^{t_0+\Delta t}}{n_{aij}^{t_0}} - \frac{\Delta t}{n_{aij}^{t_0}}\left(\sum_{kl \neq ij} n_{akl} P_{kl,ij} - n_{aij} \sum_{kl \neq ij} P_{ij,kl}\right.  \nonumber \\
 &+& \left. \sum_{r^+} n_{r^+,A} n_{r^+,B}n_{r^+,C} K_{r^+} \right.\nonumber\\
 &-&\left. \sum_{r^-} n_{r^-,A} n_{r^-,B}n_{r^-,C} K_{r^-} \right) - 1  = 0, \label{eqn:rates}
\end{eqnarray}
where we include the ground level, four excited states and the continuum. The radiative and collisional rates used for atomic hydrogen follow the method of \cite{sollum_thesis_1999}, and are described in Appendix \ref{sec:sollum_rates}. The rate equation describing molecular hydrogen is 
\begin{eqnarray}
f_{8} &=& \frac{n_{H_2}^{t_0+\Delta t}}{n_{H_2}^{t_0}} - \frac{\Delta t}{n_{H_2}^{t_0}}\left(\sum_{r^+} n_{r^+,A} n_{r^+,B}n_{r^+,C} K_{r^+} \right.\nonumber\\
&-& \left. \sum_{r^-} n_{r^-,A} n_{r^-,B}n_{r^-,C} K_{r^-}\right) - 1  = 0, \label{eqn:H2rates}
\end{eqnarray}
where the rates for the formation and destruction of molecular hydrogen are listed in Appendix \ref{sec:Hmol_rates}. We use 5 out of 6 of the atomic hydrogen rate equations and the rate equation of molecular hydrogen. We include the nucleus conservation equation, and discard the rate equation of the level with the highest population. In principle this method can be extended to include an arbitrary choice of atoms and molecules.

In the wake of strong chromospheric shocks, the internal energy density of the plasma can become very low. In non-equilibrium simulations, recombination is too slow for ionisation energy to be released as heat. Very low temperatures may lead to the EoS and opacity tables becoming inaccurate, or to poor convergence of the solver when the $H_2$ fraction becomes dominant. It is therefore necessary to include additional mechanisms to prevent temperatures becoming too low. We include three mechanisms. Firstly, an additional time-step constraint is included. The time-step $\Delta t$ is limited such that $|Q_\mathrm{rad}| \Delta t/E_{\mathrm{int}}\leq 0.25$, $Q_{rad}$ is the total radiative cooling/heating. This damps large decreases in energy due to radiative cooling.  Secondly, a minimum temperature threshold is set. Rather than including a parameterised heating term, as in \citet{leenaarts_2011_minimum}, a temperature floor is implemented. For a minimum temperature $T_{\mathrm{min}}$, a minimum value $E_{\mathrm{min}}$ for the instantaneous parts of the internal energy equation is calculated. This includes the kinetic $n k_{\mathrm{B}} T_{\mathrm{min}}$ term, an increase in the ionisation and excitation energies of non-hydrogen atoms and the dissociation of molecules treated in chemical equilibrium. The internal energy is then limited to this minimum value after a call to the EoS routine. In order to match the simulation of \citet{carlsson_2016_public} we use $T_{\mathrm{min}}=2500~\mathrm{K}$. Finally, if the populations will not converge to the required tolerance, due to the temperature dropping below a threshold value of $1000~\mathrm{K}$ during a solver call, we allow a small amount of $H^+$ to be recombined to ensure convergence. This allows faster convergence of grid points which will be limited anyway by the $2500~\mathrm{K}$ floor.

\subsection{Scattering Multi-group  Radiation Transfer}\label{sec:multigroup}

The MURaM code uses a multigroup method \citep{Nordlund_1982_theOG} to accurately and efficiently compute the frequency-dependent photospheric radiation field \citep{voegler_2004_approximations}. To accurately simulate the low-chromosphere the treatment of radiation is extended to include a scattering term. We follow the prescription of \citet{skartlien_2000_multigroup} and the extension of \citet{hayek_2010_radiative} to short-characteristics, see also \citet{collet_2011_three}. Calculation of the radiation field requires solution of the time-independent radiative transfer equation
%
\begin{equation}
\label{eqn:rte_equation}
\frac{d I_{\nu}}{d \tau_{\nu}} = S_{\nu} - I_{\nu},
\end{equation}
 where $\nu$ is the frequency, $S_{\nu}$ the source function, $I_{\nu}$ is the specific intensity, and $d\tau_{\nu} = \chi_{\nu} ds$ is the optical thickness over a path length $ds$, where $\chi_{\nu}$ is the plasma opacity. Eqn. \ref{eqn:rte_equation} must be solved for a number of ray directions $\mathbf{\hat{n}}$. The source function  has been expanded to include a scattering term,
\begin{equation}
\label{eqn:source_function}
S_{\nu} = \left( 1 - \epsilon_{\nu} \right) J_{\nu} + \epsilon B_{\nu},
\end{equation}
where $\epsilon_{\nu}$  is the photon destruction probability, $B_{\nu}$ the LTE (Planck) source function, and the mean intensity $J_{\nu}$ is calculated as the integral over all angles
\begin{equation}
\label{eqn:avg_intensity}
J_{\nu} = \frac{1}{4\pi} \int_{4\pi} I_{\nu}\left(\mathbf{\hat{n}}\right) d \omega.
\end{equation}
The type A quadrature of \citet{carlson_1963_quadrature}, including three points per quadrant, is used to perform the angular integration. The radiative energy flux $\mathbf{F_{\nu}}$ is
\begin{equation}
\label{eqn:avg_flux}
\mathbf{F_{\nu}} = \int_{4\pi} I_{\nu}\left(\mathbf{\hat{n}}\right) \mathbf{\hat{n}} d \omega.
\end{equation}

 For the formal solution to the radiative transfer Eqn. \ref{eqn:rte_equation} we use a short characteristics scheme with linear interpolation. The intensity at a given point $(O)$ is calculated as
\begin{equation}
\label{eqn:formal_solution}
I_O = I_U e^{-\delta \tau_U} + \Phi_U S_U + \Phi_O S_O,
\end{equation}
where $(U)$ is the upwind point, $\tau_U$ is the optical distance on the segment $UO$ and the $\Phi$ quantities are the weights. The formal solution, Eqns. \ref{eqn:source_function} \& \ref{eqn:formal_solution}, can be written as
\begin{equation}
\label{eqn:lambda_operator}
J_{\nu} = \Lambda\left[ S_{\nu} \right] + \mathcal{J_{\nu}},
\end{equation}
where $\mathcal{J_{\nu}}$ represents the transmitted contribution to $J_{\nu}$ due to the given incident radiation at the boundaries of the computational domain, and $\Lambda$ is the angle-averaged Lambda operator. 

Direct solution of Eqn. \ref{eqn:lambda_operator} is expensive, and simply updating Eqn. \ref{eqn:source_function} with the new $J_{\nu}$ leads to slow convergence. We employ the Approximate Lambda Iteration (ALI) method \citep{cannon_1973_frequency}, and the diagonal operator as the approximate operator. 
%
%
%
%
%
%
The scheme is iterated until a tolerance of $10^{-4}$ is reached on the relative {\bf correction} of the source function. Once the source function has converged the radiative cooling/heating $Q_{\nu}$ is calculated from two equivalent expressions
\begin{equation}
\label{eqn:Q_rad}
Q_{RT} = - \int_{\nu} \left( \nabla \cdot \mathbf{ F_{\nu}}\right) d \nu = 4 \pi \int_{\nu} \chi_{\nu} \left(J_{\nu} - S_{\nu}\right) d \nu.
\end{equation}

The multi-group scheme is used to simplify the frequency spectrum into a number of subsets $j$, known as bands. Instead of detailed calculations incorporating $10^3 - 10^5$ frequency points, many of the important processes for a radiative MHD magneto-convection simulation, such as line blanketing, can be captured by using as few as $4-5$ bands \citep{voegler_2004_approximations}. To calculate the radiation field we require three band-integrated quantities, the extinction coefficient $\chi_{j}$, the mean scattering albedo $\left(1-\epsilon\right)_{j}$ and the band-integrated emissivity $\left(\epsilon B\right)_{j}$. The binning process is discussed in detail in Appendix \ref{sec:scattering_opacities}. The integral over wavelength in Eqn \ref{eqn:Q_rad} is then solved as
\begin{eqnarray}
\label{eqn:Q_rad_sum}
Q_{RT} &=& - \sum_j \left[ \left( \nabla \cdot \mathbf{ F_j}\right) \left(1-e^{-\tau_j/\tau_0}\right) \right.\nonumber\\
&+& \left. 4 \pi \chi_j \left(J_j - S_j\right) e^{-\tau_j/\tau_0}\right]
\end{eqnarray}
where $\tau_j$ is the band mean optical depth, $\tau_0 = 0.1$ and the term $e^{-\tau_j/\tau_0}$ provides a transition between the radiative energy and flux-divergence form of the equation. This prevents numerical round-off errors in the optically thick regime where $J \approx S$, which are amplified as $\chi$ grows exponentially with depth \citep{Bruls_1999_radiativeheating}. 

\subsection{Radiative Cooling/Heating}\label{sec:radiation}

To perform radiation MHD simulations from the convection zone to the corona a radiation scheme is required which can accurately model a range of physical regimes. These include the deep interior, which can be treated using the diffusion approximation. The photosphere where spectral line formation becomes important, requiring a 3D multi-group approach (Sect. \ref{sec:multigroup}). Finally, in the chromosphere the radiation transport must include NLTE effects, such as PRD and scattering. Detailed NLTE radiation transfer is too computationally expensive to be performed in a 3D time-dependent simulation. To include accurate and fast radiative cooling/heating in the chromosphere and corona, we use the pre-tabulated radiative losses calculated by \citet{carlsson_2012_approximations}.

The prescription of \citet{carlsson_2012_approximations} consists of two parts; NLTE line losses in the chromosphere from hydrogen, calcium and magnesium, and optically thin coronal losses. We use the overlap interval approach of \cite{rempel_2017_extension}. In the original implementation the overlap interval is calculated only in the vertical direction. In this work we take the isotropic average of the overlap interval in all three directions to better model the cooling around the irregularly shaped transition region.

In the chromosphere the most important spectral lines and continuua are the Lyman-$\alpha$, H-$\alpha$ and the Lyman continuum of hydrogen, the Mg~{\sc ii} h \& k and the Ca~{\sc ii} H \& K lines. These lines are modelled using a simplified description of the heating/losses, for element $a$ in ionisation stage $i$;
\begin{equation}
\label{eqn:overlaplinelosses}
Q_X = -f(T) q_{a,i}(\tau) n_{a,\mathrm{tot}} n_{\mathrm{e}} ,
\end{equation}
 where $f(T) = L_{a,i}(T) F_{a,i}(T)$. Here three pre-tabulated quantities\footnote{The tables for the chromospheric lines are available as part of an IRIS data release \url{http://iris.lmsal.com/bf/code.tar.bz2}.} are used; $L_{a,i}(T)$ is the optically thin radiative loss function, $F_{a,i}(T)$ is the fraction of element $a$ in ionisation stage $i$, and $q_{a,i}(\tau)$ is the escape probability as a function of some optical depth proxy $\tau$. For the escape probability of calcium and magnesium the column mass is used as a proxy for $\tau$, and for hydrogen the neutral hydrogen column density is used. The optically thin radiative losses $Q_{\mathrm{thin}}$ are calculated as;
\begin{equation}
\label{eqn:thinlosses}
Q_{\mathrm{thin}} = -  \Lambda(T) n_{\mathrm{H,tot}} n_{\mathrm{e}},
\end{equation}
where $\Lambda(T)$ is given as a table in terms of temperature.

Additionally, we include the back-heating of chromospheric plasma ($Q_{back}$). This is performed using the 3D radiation transport scheme, following \citet{carlsson_2012_approximations}. The emissivity is given in terms of the optically thin coronal losses given by:
\begin{equation}
\eta_{back} = -\frac{Q_{\mathrm{thin}}}{4 \pi},
\end{equation}
and the opacity at the ionisation edge of helium is used;
\begin{equation}
\chi_{\mathrm{back}} = \alpha \frac{n_{\mathrm{He\;I}}}{n_{\mathrm{He}}}\left(T,p_e \right) \frac{n_{\mathrm{He}}}{\rho},
\end{equation}
 where $\alpha$ is the opacity at the ionisation edge of helium, $n_{\mathrm{He\;I}} /n_{\mathrm{He}}$ is the neutral helium fraction and is pre-tabulated in LTE in terms of electron pressure and temperature, and $n_{\mathrm{He}}/\rho$ is the number of particles of helium per gram of stellar material.

The full radiative cooling/heating prescription is then:
\begin{equation}
Q_{\mathrm{rad}} = Q_{\mathrm{RT}} + Q_{\mathrm{H}} + Q_{\mathrm{Mg}} + Q_{\mathrm{Ca}} + Q_{\mathrm{thin}} + Q_{\mathrm{back}},
\end{equation}
where $Q_{\mathrm{RT}}$ is the heating/cooling from the multi-group radiative transport scheme described in Sect. \ref{sec:multigroup} above. To prevent over-cooling from both the LTE and NLTE losses in the upper chromosphere, the cooling in each radiation band is switched off based on the band-averaged optical depth $\tau$. This is performed using a function of the form $\tau^2/\left(\tau^2 + \tau_{\mathrm{cutoff}}^2 \right)$. A value of $\tau_{\mathrm{cutoff}}=1.0\times 10^{-4}$ is used in these simulations. The multi-group RT scheme groups frequencies into bands. Which band a particular frequency $\nu$ goes into depends on the height where $\tau_{\nu} = 1$. We use as a reference $\tau_{500}$, the optical depth at $500~\mathrm{nm}$.  Previous work has found 4-bands are sufficient to capture back-heating and line-blanketing in the photosphere and temperature minimum \citep{voegler_2005_muramcode}. We use a 4-band setup similar to \citet{carlsson_2016_public}, with boundaries at the heights where $\tau_{500} = 10^{-1/2}$, $10^{-3/2}$, and $10^{-5/2}$.

\subsection{MHD Equations}\label{sec:MHD}

The set of equations solved by the MURaM code, are
\begin{eqnarray}
\pd{\rho}{t} &=& -\nabla \cdot \left(\rho \mathbf{v} \right) \\
\pd{\rho \mathbf{v}}{t} &=& -\nabla \cdot \left(\rho \mathbf{v} \mathbf{v}\right) - \nabla p + \rho \mathbf{g} + \mathbf{F}_{\mathrm{L}} + \mathbf{F}_{\mathrm{SR}} \\
\pd{E_{\mathrm{HD}}}{t} &=&  -\nabla \cdot \left[ \mathbf{v} \left(E_{HD} + p\right) + q\mathbf{\hat{b}}\right]+\rho \mathbf{v} \cdot \mathbf{g} + \mathbf{v} \cdot \mathbf{F_L} \nonumber \\
&+& \mathbf{v} \cdot \mathbf{F_{\mathrm{SR}}} + Q_\mathrm{rad} + Q_{\mathrm{res}} \\
\pd{\mathbf{B}}{t} &=& \nabla \times \left(\mathbf{v} \times \mathbf{B}\right)
\end{eqnarray}
where $\rho$ is the plasma density, $\mathbf{v}$ the velocity vector, $p$ the gas pressure, $\mathbf{g}$ the gravitational acceleration, $\mathbf{F}_{\mathrm{L}}$ the Lorentz force, $\mathbf{F}_{\mathrm{SR}}$ the semi-relativistic (Boris) correction, $E_{\mathrm{HD}}$ the hydrodynamic energy, q the Spitzer heat flux, $\mathbf{\hat{b}}=\mathbf{B}/|\mathbf{B}|$ the unit vector in the direction of the magnetic field vector $\mathbf{B}$, $Q_\mathrm{rad}$ the radiative cooling/heating, $Q_{\mathrm{res}}$ the resistive heating, and $T$ the gas temperature. Additional diffusive terms are applied to each equation, based on the scheme described in Sect. \ref{subsec:diffusion},

The various radiative heating and cooling effects are given by $Q_\mathrm{rad}$ and described in detail in Section $\ref{sec:radiation}$ above. The hydrodynamic energy $E_{\mathrm{HD}}=E_{\mathrm{int}} + \frac{1}{2} \rho \mathbf{v}^2$ is used instead of the total energy. This prevents numerical errors in calculating $E_{\mathrm{int}}$ from $E_{\mathrm{tot}}$ in low-$\beta$ regions where the magnetic energy dominates the total energy. To conserve the total energy the heating from the diffusion scheme is then added as $Q_{\mathrm{res}}$. The Spitzer heat flux $q$ is solved using the hyperbolic method, see \cite{rempel_2017_extension} for a full derivation,
\begin{equation}
\pd{q}{t} = \frac{1}{\tau}\left(-f_{\mathrm{\mathrm{sat}}}\sigma T^{\frac{5}{2}}\left(\mathbf{\hat{b}}\cdot\nabla\right)T-q\right)\\
\end{equation}

where $\sigma=1.1\times 10^{-6}~\mathrm{erg\;cm^{-1}\;s^{-1}\;K^{-7/2}}$ is the constant of Spitzer heat conductivity, $f_{\mathrm{sat}}$ controls the saturation of thermal conduction and $\tau$, which is used to control the transition between parabolic and hyperbolic solutions to the heat conduction equation, has the form
\begin{equation}
\tau = \left(f_{\mathrm{CFL}}\frac{\Delta x_{\mathrm{min}}}{\Delta t} - |v|\right)^{-2} \frac{f_{\mathrm{sat}} \sigma T^{7/2}}{E_{\mathrm{int}}},
\end{equation}
where $f_{\mathrm{CFL}} \frac{\Delta x_{\mathrm{min}}}{\Delta t} - |v|$ is used as a maximum propagation speed in order to avoid violations of the Courant–Friedrichs–Lewy (CFL) condition. The first term is chosen so that the maximum wave speed of the hyperbolic heat conduction is comparable to the maximum MHD wave speed with the limited Alfv\'en Velocity $(c = f_{\mathrm{CFL}} \frac{\Delta x_{\mathrm{min}}}{\Delta t})$, in terms of the minimum spatial grid-scale $x_{\mathrm{min}}$ and time-step $\Delta t$. In order to explicitly integrate the system of equations we set a lower limit on $\tau$ of $4 \Delta t$.

The treatment of the Lorentz force $\mathbf{F}_{\mathrm{L}}$ follows that of \citet{rempel_2017_extension}, 
\begin{eqnarray}
\mathbf{F}_{\mathrm{L}} &=& \frac{f_{A}}{4 \pi} \nabla \cdot \left( \mathbf{B} \mathbf{B} - \frac{1}{2}\mathbf{I}\mathbf{B}\right) \nonumber\\
&+& \frac{1-f_A}{4 \pi}  \left(\nabla \times \mathbf{B} \right) \times \mathbf{B},
\end{eqnarray}
where $f_A=\frac{1}{\sqrt{1+\left(v_a/c_{\mathrm{max}}\right)^4}} $ is the Alfv\'en limit factor in terms of the reduced speed of light $c_{\mathrm{max}}$ and Alfv\'en speed $v_a$. This form is used to have a sharper transition between the limited and non-limited regime, the semi-relativistic form would give $1/(1+(v_a/c_{\mathrm{max}})^2)$. The reduction of the Alfv\'en velocity is achieved through a semi-relativistic treatment with reduced speed of light (Boris correction), which can be implemented through a projection operator in the momentum equation \citep{gombosi_2002_semirelativistic_MHD,rempel_2017_extension} by adding the force term $\mathbf{F}_{\mathrm{SR}}$ given by

\begin{eqnarray}
\mathbf{F}_{\mathrm{SR}} &=& - \left(1-f_A\right) \left[\mathcal{I} -\mathbf{\hat{b}} \mathbf{\hat{b}}\right] \nonumber \\
 && \left(-\rho \left(\mathbf{v}\cdot \nabla\right)\mathbf{v} - \nabla p + \rho \mathbf{g} + \mathbf{F_L}+ \nabla \cdot \boldsymbol{\tau}^{\mathrm{diff}}\right).
\end{eqnarray}
 The components of the numerical viscous stress tensor ($\boldsymbol{\tau}^{\mathrm{diff}}$) are calculated
\begin{equation}
\left(\nabla \cdot \boldsymbol{\tau}^{\mathrm{diff}}\right)_i =\sum_{j=x,y,z} \frac{1}{\Delta_j}\left(f^{j+1/2}_{v_i}-f^{j-1/2}_{v_i}\right) ,
\end{equation}
where $\Delta_j$ is the grid resolution, and $f^j_{v_i}$ are the diffusive fluxes of velocity component $i$ in the $j$ direction, calculated using Eqn. \ref{eqn:diff_flux}.

\section{Simulation Setup} \label{sec:setup}

We present results of a simulation continued from the publicly available Bifrost simulation \citep{carlsson_2016_public}. The initial condition consists of a bipolar magnetic field region modelling an enhanced network magnetic field. The simulation has been run in non-equilibrium for 3850 seconds, starting from the first publicly available snapshot "285".

\begin{figure*}[htp]
\centering
\includegraphics[width=16cm]{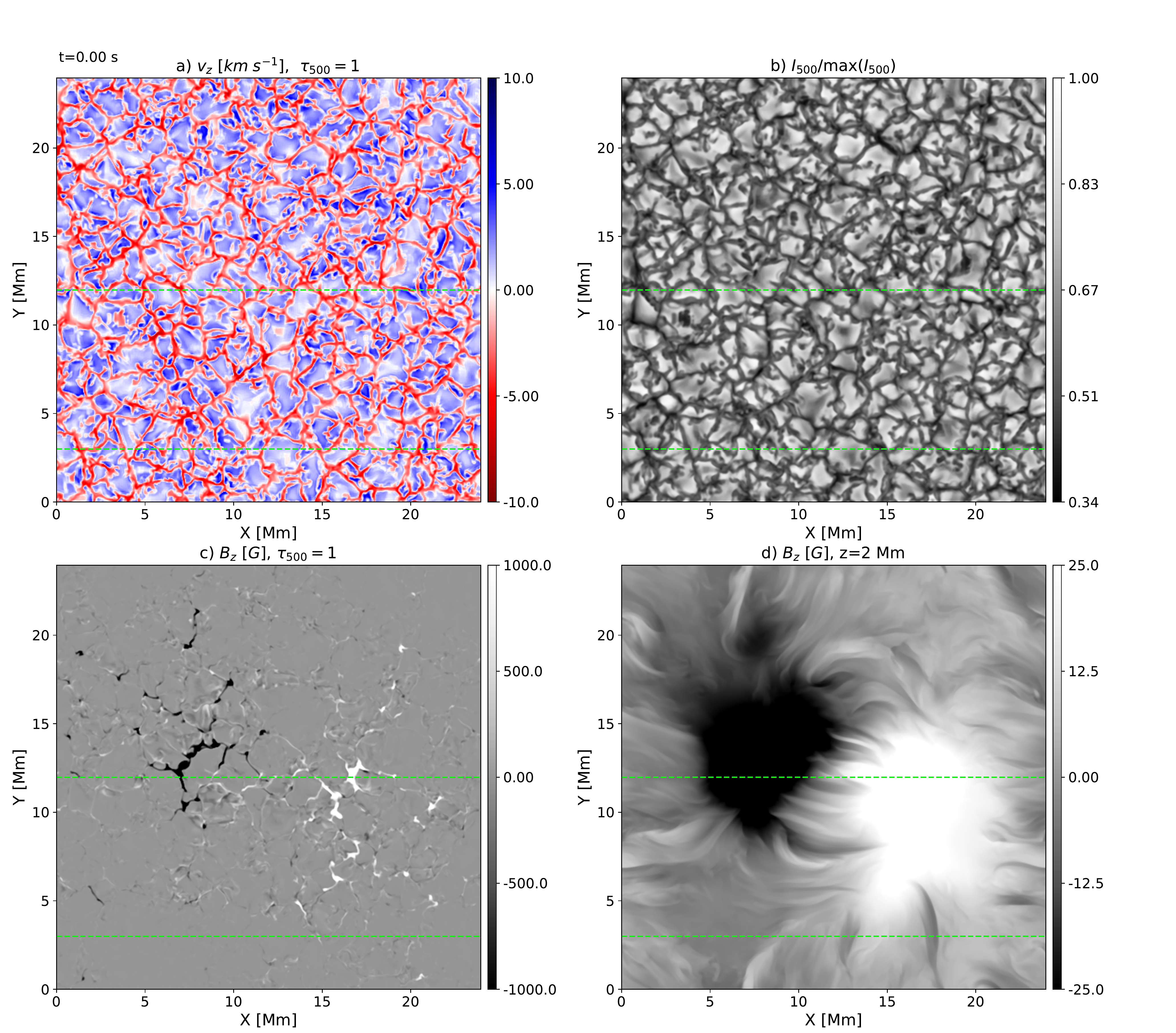}
\caption{A snapshot of the simulation after 3850s, showing: panel a) The vertical velocity at the photosphere, b) the normalised intensity at $500~\mathrm{nm}$, panel c) the vertical magnetic field at the photosphere, and panel d) the vertical magnetic field at a height of $2~\mathrm{Mm}$ in the chromosphere. The green dashed lines show the slices taken in Fig. \ref{fig:chromo_slices}. Panels c) and d) have been saturated in order to show the fine structure of the magnetic field. Slices at the photosphere are taken at the contour where $\tau_{500}=1$.}
\label{fig:initial_snapshot}
\end{figure*}

One difference between the two simulations comes from the use of a stretched grid in the Bifrost code. This must be interpolated onto a constantly spaced vertical grid suitable for the MURaM code. This is performed using log-linear interpolations for density, energy, electron and population numbers, and linear interpolations for velocities and magnetic fields.

 A different equation of state is used by the Bifrost and MURaM codes. The Bifrost code uses a non-ideal EoS based on the free-energy minimisation method \citep{gustafsson_1975_grid}. We calculate an equation of state using the same abundances. There remain inconsistencies between the two EoS's, coming from the non-ideal formulation used by Bifrost, and differences in partition functions for atoms and molecules. The internal energy (Eqn \ref{eqn:NE_energy}) was recalculated with the EoS described in Section \ref{subsec:EoS}, using the hydrogen population levels, temperature and electron number densities of the original simulation. The relative difference in the internal energy has a rms value of $3 \times 10^{-3}$, with a maximum of $6 \times 10^{-2}$ near the transition region.

The resulting simulation has $504 \times 504 \times 840$ grid-points, spanning $24\Mm$ in the horizontal direction and $16.8\Mm$ in the vertical, with the lower boundary at $-2.44\Mm$, where 0 is the averaged solar surface. This corresponds to a horizontal resolution of $47.6\km$ and a vertical resolution of $20\km$. The diffusion scheme is described in Section \ref{subsec:diffusion}, and a PDM coefficient of $C_{\mathrm{pdm}}=2$ is used for both the diffusion scheme and the population advection. The viscous and resistive heating is determined from the momentum and magnetic field fluxes as described in \citep{rempel_2017_extension}. We enforce a minimum temperature of $2500\;$K to prevent over-cooling in the post-shock rarefactions. The upper boundary condition imposes a potential field and is open to outflows, but closed to inflows. The lower boundary condition is the Open Symmetric-field (OSb) condition described by \citet{rempel_2014_numerical}. 

 We limit the Alfv\'en speed through the use of the Boris correction (semi-relativistic MHD with an artificially reduced speed of light), in combination with the dynamic limiting scheme described by \citet{rempel_2017_extension}. The maximum speed of light used in the Boris correction is calculated $c_{\mathrm{max}} = \max\left(2 c_{\mathrm{s},\;\mathrm{max}},3 v_{max}\right)$ in terms of the maximum sound speed ($c_{\mathrm{s},\;\mathrm{max}}$) and velocity ($v_{max}$) in the simulation domain. In addition, we impose a dynamic ceiling on velocity ($v_{\mathrm{max}}$) and internal energy ($\epsilon_{\mathrm{max}}$) in order to prevent extreme values, only realised in a few grid points, dominating the simulation time step. The value of the ceiling $v_{\mathrm{max}}$ is chosen dynamically so that it affects fewer than one in one million grid points.
If more than $9.4\times 10^{-7}$ of the simulation grid-points are above 0.95 of the limited  $v_{\mathrm{max}}$ (or $\epsilon_{\mathrm{max}}$), then the value is increased by 1\%. If fewer than $4.7 \times 10^{-7}$ points are above 0.95 of the limit then is lowered by 1\%. The simulation presented has a maximum velocity of $300-400~\mathrm{km\;s}^{-1}$. The maximum speed of light used in the Boris correction is then calculated $c_{\mathrm{max}} = \max\left(2 c_{\mathrm{s,max}}, 3 v_{\mathrm{max}} \right)$. We do not allow the speed of light in the box to decrease below $2000~\mathrm{km\;s}^{-1}$. The chosen limits on velocity and speed of light in the box ensure a minimal effect on the chromospheric structure and dynamics. The impact of the choice of maximum speed of light has been studied for strong field active region simulations \citep{rempel_2017_extension,Warnecke_2020_heat_and_Boris}. The simulation presented in this work has a significantly weaker magnetic field than the active region simulations, and far fewer grid-points will require limiting.

When the simulation {\bf was} started 
the potential field upper boundary condition causes the coronal field to become more vertical. The transition region then produces a large transient. Differences resulting from the interpolation, and slight differences in the way the EoS  was constructed, likely contribute to this transient. To reduce the timescale of the transient the diffusion was temporarily increased on velocities over $100~\mathrm{km\;s}^{-1}$ at the start of the simulation. The simulation was run until this transient passed and the RMS velocity stabilised, which took about 600 seconds. The additional diffusion and damping is then slowly removed and the simulation is run for an hour until the corona reheated. 

\begin{figure*}[htp]
\centering
\includegraphics[width=16cm]{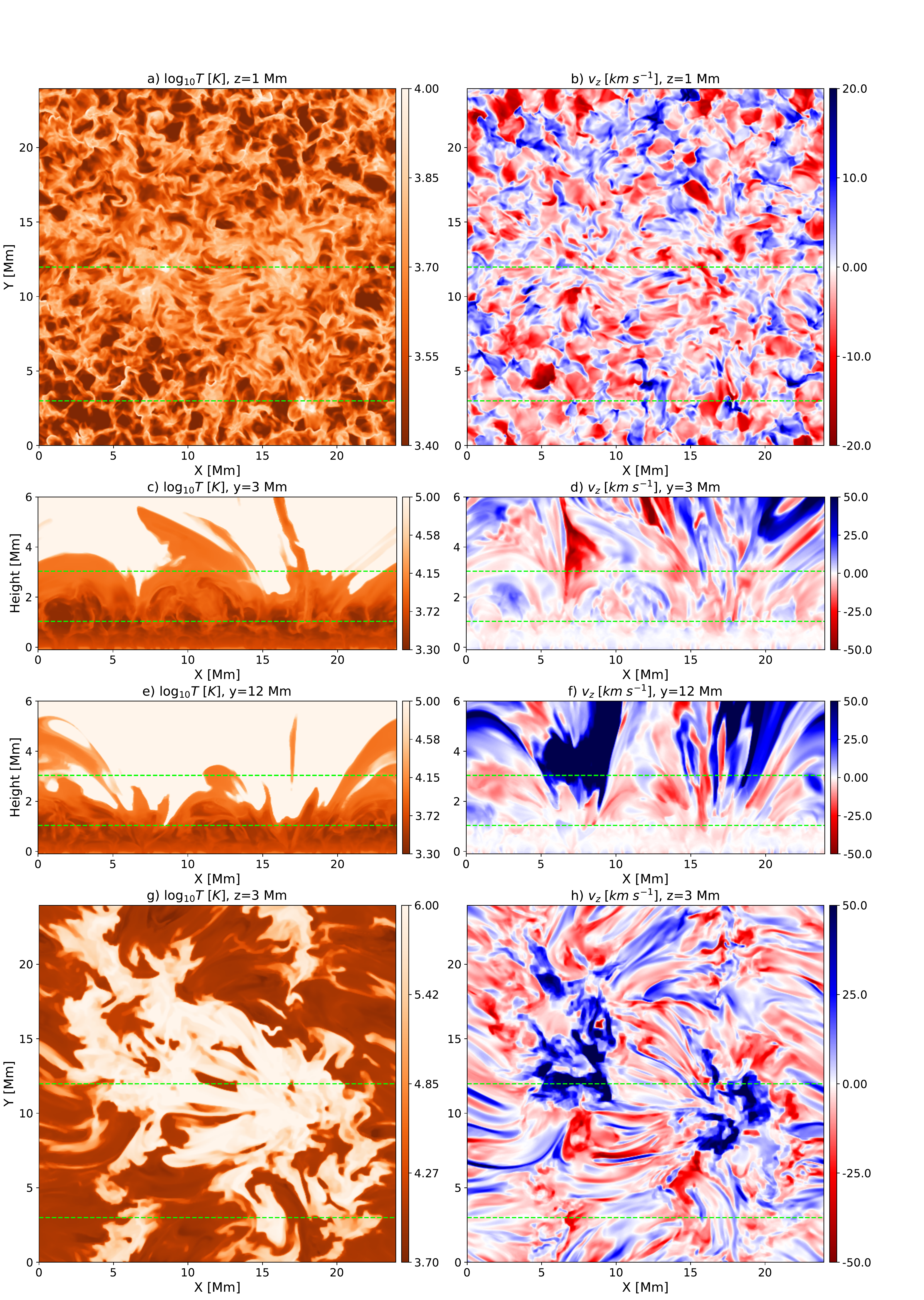}
\caption{Slices through the chromosphere after 3850s of the simulation, showing temperature in the left column and vertical velocity in the right column. Slices in the horizontal plane are taken through the low chromosphere (panels a \& b) and the upper chromosphere (panels g \& h). The middle rows show vertical slices through a quieter inter-network region (panels c \& d) and through the \textbf{centre} of the network field (panels e \& f). Animation available online.}
\label{fig:chromo_slices}
\end{figure*}

\begin{figure*}[htp]
\centering
\includegraphics[width=16cm]{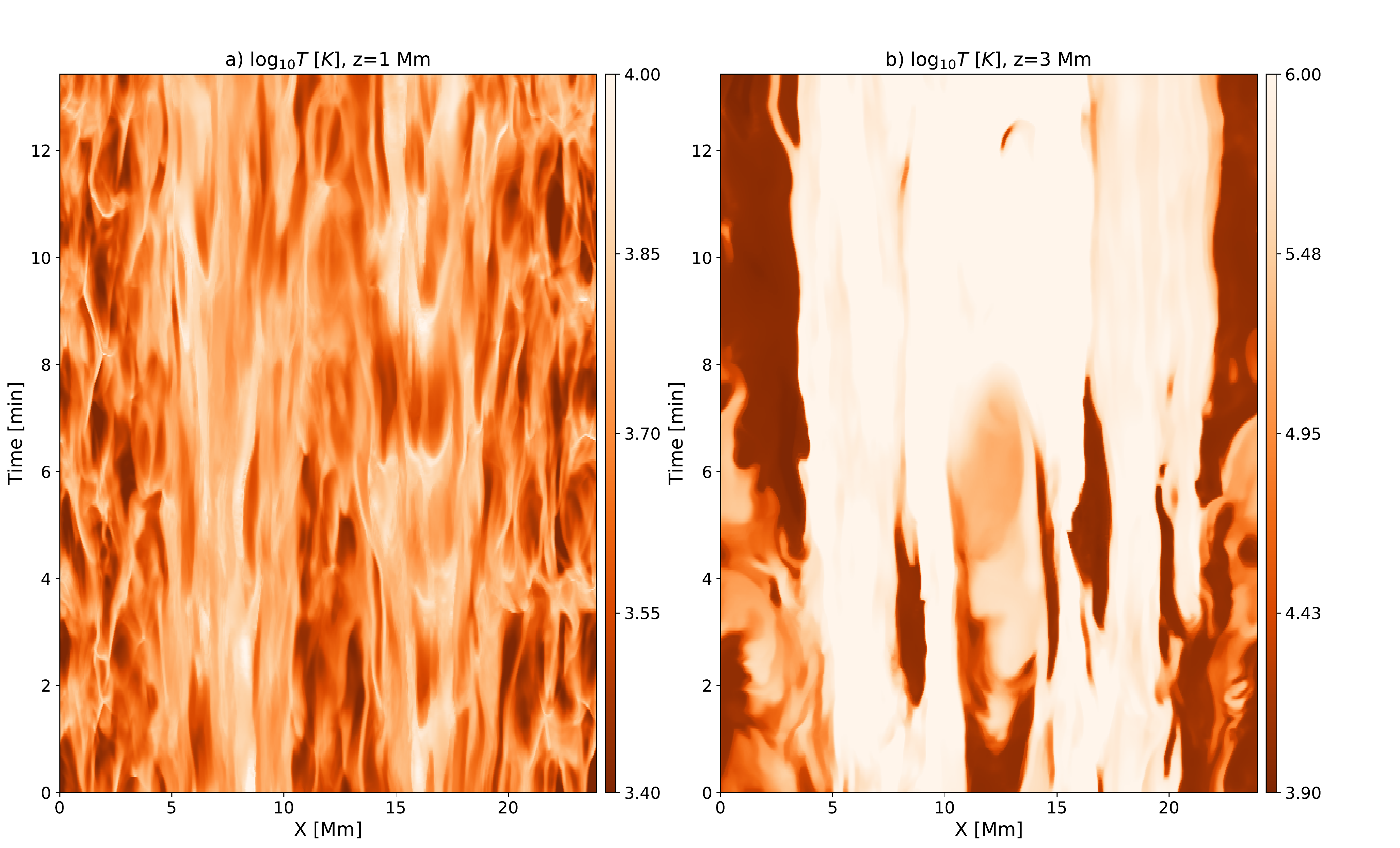}
\caption{A time distance diagram of temperature spanning 14 minutes of the simulation. The slice is taken at $y=12~\mathrm{Mm}$, and at two heights $z=1\Mm$ (panel a) and $z=2\Mm$ (panel b).}
\label{fig:chromo_tds}
\end{figure*}

This simulation cannot be directly compared to the original Bifrost public snapshot due to the differences resulting from the large transient and the lower viscosity and resistivity. The resulting model is shown in Fig. \ref{fig:initial_snapshot}. The photosphere shows a bipolar enhanced network regions contain strong field concentrations of $1-3\;$kG. The magnetic field in the chromosphere (panel d) is dominated by the large-scale bipolar fields, with finely structured strands, including regions of opposite polarity.

The chromospheric dynamics are shown in Fig. \ref{fig:chromo_slices} \footnote{\color{blue} \href{https://drive.google.com/file/d/1Xlc_N0fzzHrN9jIuxdeVwe0FKGxs2O8G/view?usp=sharing}{Animation Available online}}, in the mid chromosphere (panel a \& b, $z=1~\mathrm{Mm}$) is dominated by shocks. The shock-fronts show velocities over $20~\mathrm{km\;s^{-1}}$ in the inter-network regions. Above the strong network fields the shocks are suppressed and the dynamics follows the magnetic field. The temperatures range from the minimum value of $2.5~\mathrm{kK}$ in the shock rarefactions to above $10~\mathrm{kK}$ in the shock fronts. Above the network fields the temperatures are higher, reaching nearly $100~\mathrm{kK}$ in regions where the transition region is depressed. The chromospheric velocity field at $2~\mathrm{Mm}$ shows strong shocks, with velocities above $25~\mathrm{km\;s}^{-1}$. At $3~\mathrm{Mm}$ loops are seen between the bipolar fields, while the inter-network regions show the shock canopy with temperatures around 1 kK. 

The time-evolution of the chromospheric shocks is shown in a time-distance diagram in Fig. \ref{fig:chromo_tds}. In the low-chromosphere (panel a) the quiet regions show shock fronts with a period of $\approx 5~\mathrm{min}$. In the network field regions periodic brightenings are seen at a $x=16~\mathrm{Mm}$ with a period of $\approx 3~\mathrm{min}$. Due to the treatment of helium in LTE, a large fraction of the plasma at $3\Mm$ height plasma sits at around $10~\mathrm{kK}$, the preferred temperature of the first ionisation stage of helium.

\section{Radiative Cooling and Heating} \label{sec:results_rad}

\begin{figure*}[htp]
\centering
\includegraphics[width=16cm]{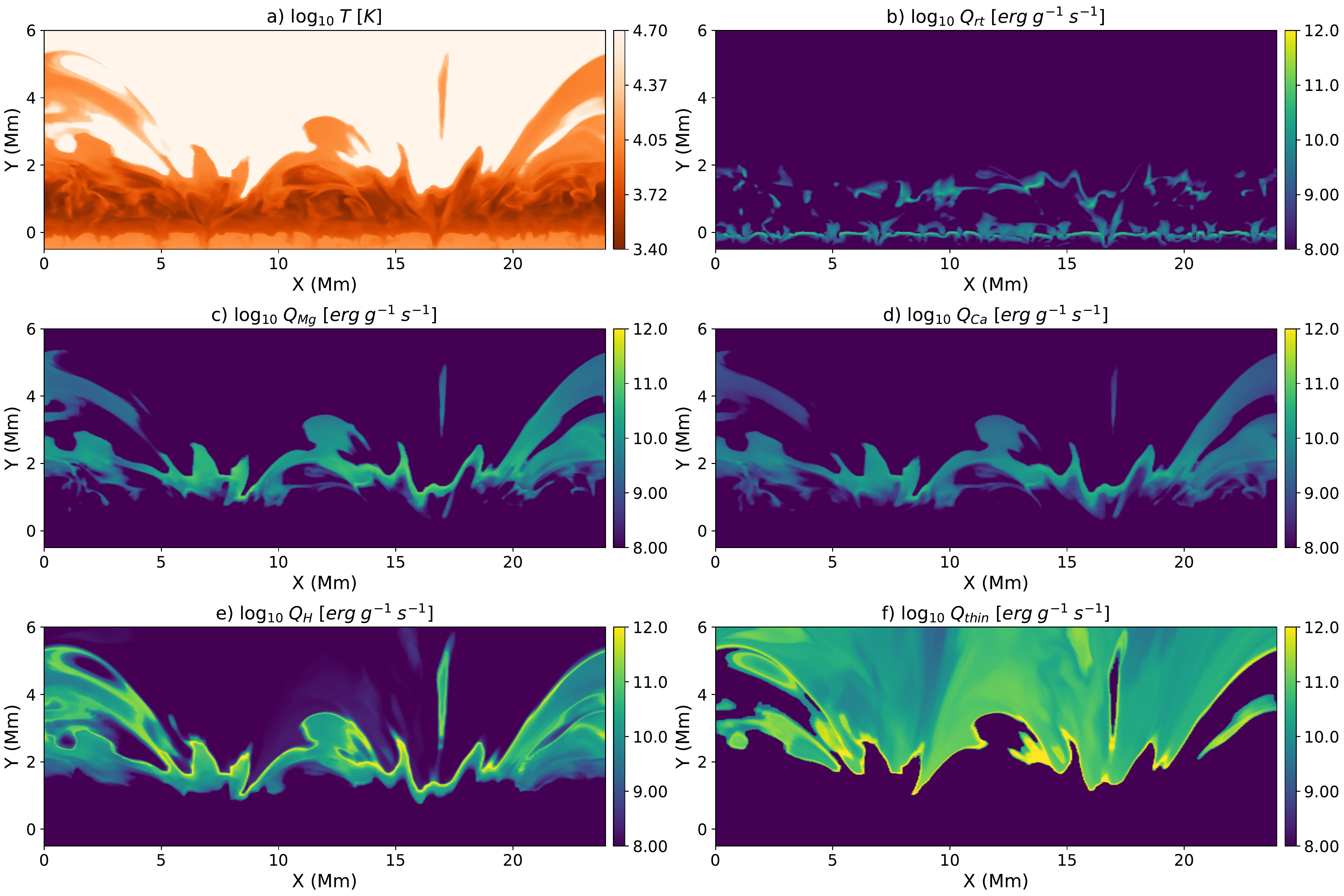}
\caption{Radiative cooling in the photosphere and chromosphere for a slice through the simulation. Panel a) shows the temperature, b) shows the losses through the 3D multi-group radiation transport scheme, the chromospheric line losses are shown for magnesium c), calcium d) and hydrogen e), and panel f) shows the optically thin losses in the corona.}
\label{fig:radiative_cooling}
\end{figure*}

\begin{figure*}[htp]
\centering
\includegraphics[width=16cm]{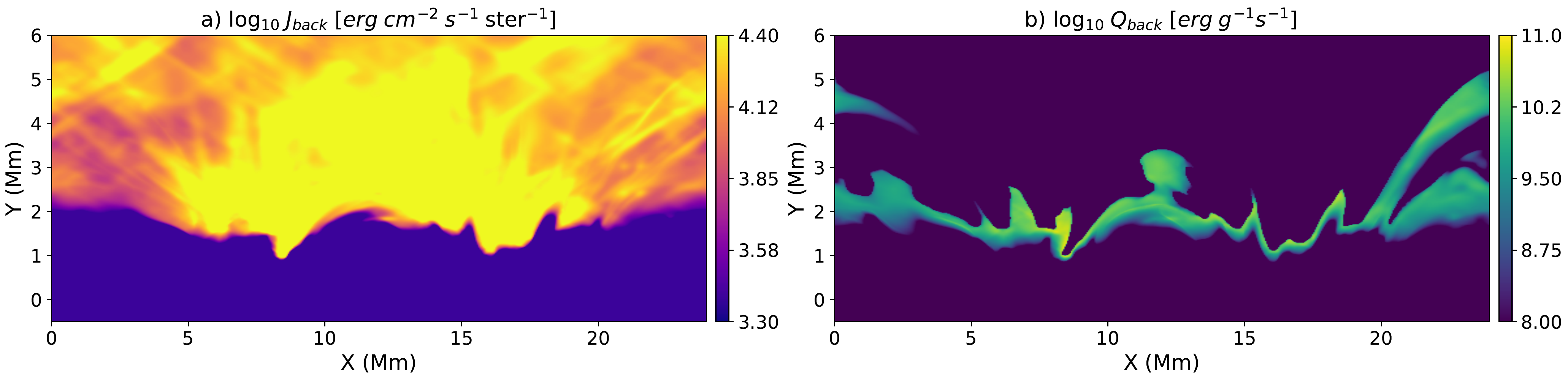}
\caption{Radiative back-heating of the upper chromosphere due to optically thin coronal losses. Panel a) shows the intensity due to the coronal losses, and panel b) shows the heating in the chromosphere.}
\label{fig:back_heating}
\end{figure*}

The radiative losses and heating in the simulation can be split into the multi-group scheme in and below the temperature minimum and low chromosphere, the chromospheric line losses, and the optically thin losses. Figure \ref{fig:radiative_cooling} shows these different components for a slice through the model. The multi-group RT scheme cools and heats the photosphere and shocks near the temperature minimum. In the low-chromosphere the calcium and magnesium losses are strongest. The hydrogen losses are strong throughout the chromosphere, dominating the upper chromosphere and peaking in the lower transition region. The optically thin losses dominate above the transition region and are strongest in a narrow region immediately above the transition region.

The prescription for Lyman-alpha and the Lyman-continuum do not provide any heating in the upper chromosphere. The prescriptions for calcium and magnesium can provide a small amount of heating when the temperature decreases below $3.5\;$kK. The EUV back-heating is strongest in the upper chromosphere, below the transition region, where neutral helium can form.  The angle averaged intensity and heating rate of the EUV bin are shown in Fig. \ref{fig:back_heating} for a slice through the simulation. High heating rates, above $10^{10}~\mathrm{erg\;g}^{-1}\mathrm{s}^{-1}$, are strongly localised near areas of the transition region where the optically thin losses are high.

\begin{figure*}[htp]
\centering
\includegraphics[width=16cm]{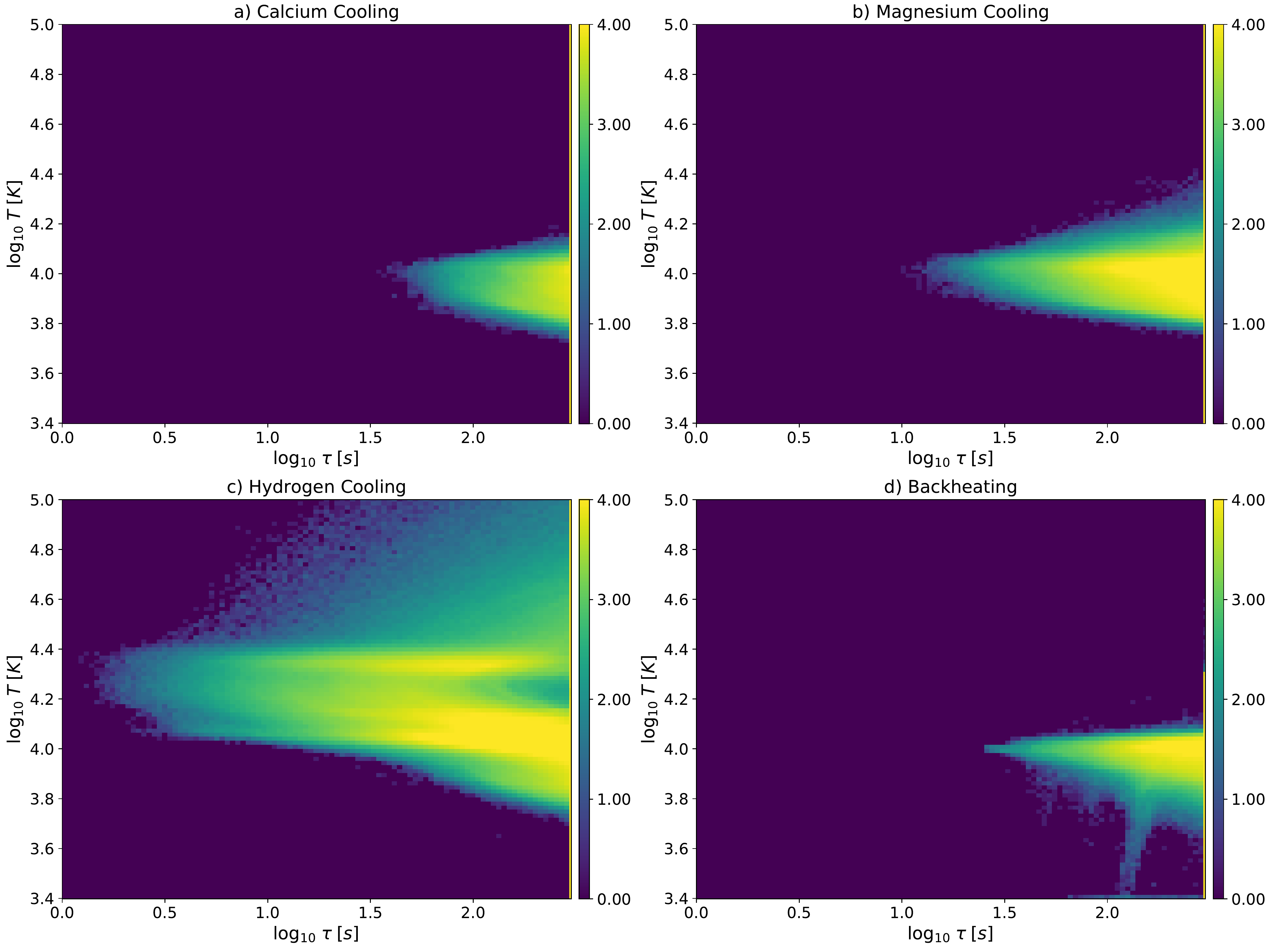}
\caption{Histograms of radiative timescales in the chromosphere with temperature, showing the recipes for a) calcium, b) magnesium, and c) hydrogen losses. Panel d) shows the back-heating from optically thin losses in the corona.}
\label{fig:timescales}
\end{figure*}

To investigate the relative importance of the different lines we plot the cooling/heating timescales in Fig. \ref{fig:timescales}. The ionisation and recombination times in the chromosphere are long, preventing fast recombination of hydrogen as the gas is cooled. We therefore calculate the timescales $\tau = E_{\mathrm{inst}}/Q$ using the terms in the internal energy that can instantaneously change $E_{\mathrm{inst}} = 3/2 N k_{\mathrm{B}} T + E_{\mathrm{nonH}} + E_{\mathrm{mol,CE}}$. This includes the microscopic kinetic energy, the ionisation of non-hydrogen species and the formation of $H^{-}$ and $H_2^{+}$ molecules in chemical equilibrium. Figure \ref{fig:timescales} shows histograms of the timescale with temperature, for the chromospheric line losses and back-heating. Calcium cooling extends to lower temperatures, affecting shocks down to the temperature minimum, but it is lower than magnesium and hydrogen through the mid-to-high chromosphere. Magnesium cooling is strongest in regions below $\approx10~kK$, and hydrogen dominates radiative losses from $\approx10kK$ to the transition region, reaching timescales lower than $10$ seconds. These results are similar to those presented in \citet{carlsson_2012_approximations}, with hydrogen dominating the cooling above the mid-chromosphere ($\approx1.5\Mm$) and being marginally lower than magnesium in the low chromosphere.

\section{Hydrogen populations} \label{sec:results_pops}

\begin{figure*}[htp]
\centering
\includegraphics[width=16cm]{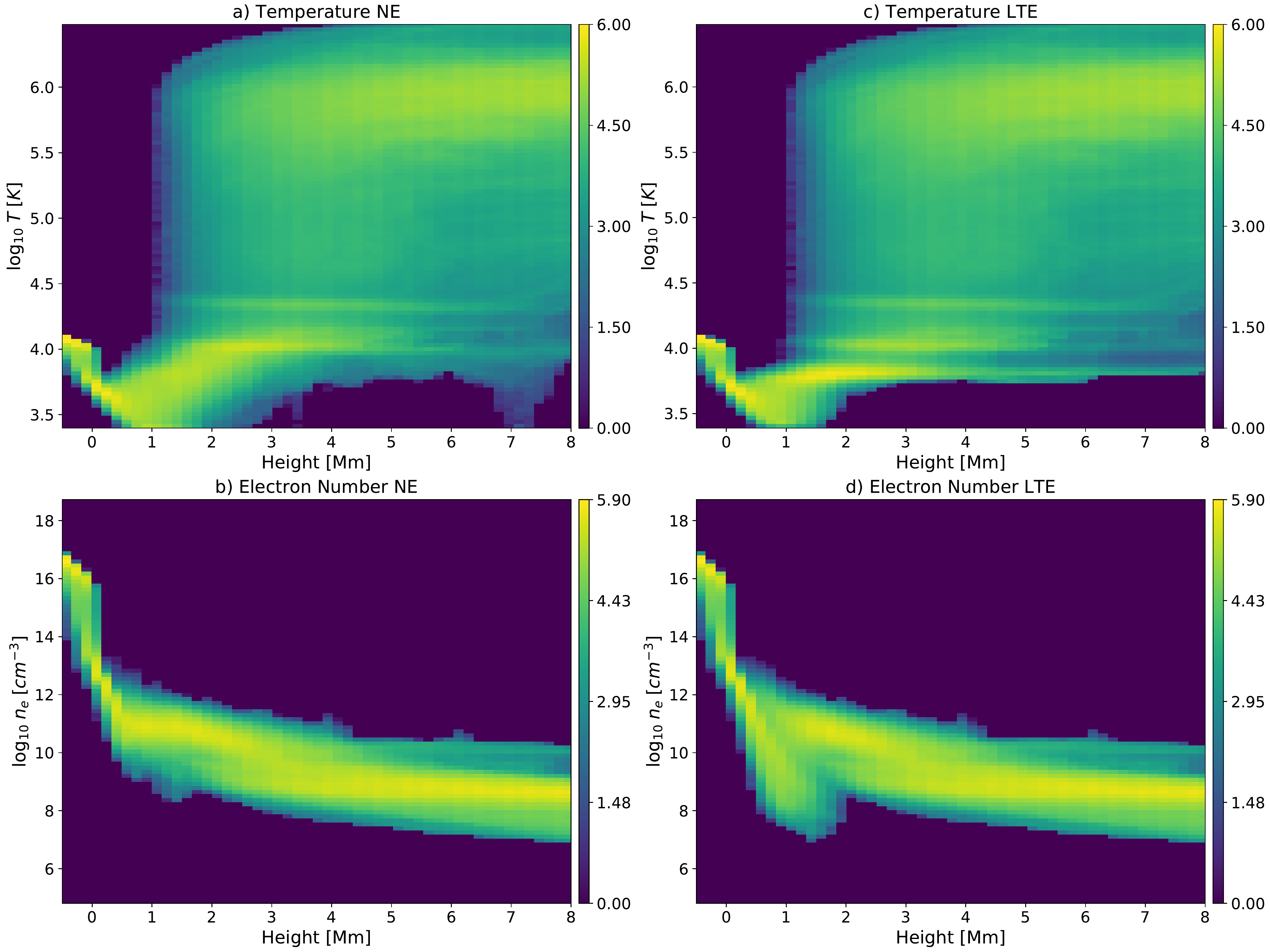}
\caption{Histograms of the temperature (top) and electron number (bottom) throughout the photosphere and chromosphere of the simulation. Comparing the non-equilibrium values (left columns) to those obtained from the pre-tabulated LTE EoS (right columns).}
\label{fig:thermodynamics_histogram}
\end{figure*}

To investigate the effects of the non-equilibrium treatment of hydrogen on the thermodynamics of the simulation we plot histograms of the temperature and electron number density in Fig. \ref{fig:thermodynamics_histogram}. The results of the non-equilibrium equation of state are compared to those calculated in LTE. The preferred temperature of hydrogen ionisation, around $8$kK is a prominent feature in the LTE results, but it is inconspicuous in the NLTE case. Two preferred temperature bands remain at $10\;$kK and $30\;$kK , caused by the first and second ionisation stages of helium, which is treated in LTE \citep{leenaarts_2011_minimum,golding_2016_NEhelium}. The spread of temperatures in the chromosphere is wider due to the long recombination timescales prevent hydrogen ionisation from buffering the temperature fluctuations. The higher ionisation fraction is also seen in the electron number density, which is higher than LTE in the low-to-mid chromosphere.

\begin{figure*}[htp]
\centering
\includegraphics[width=17cm]{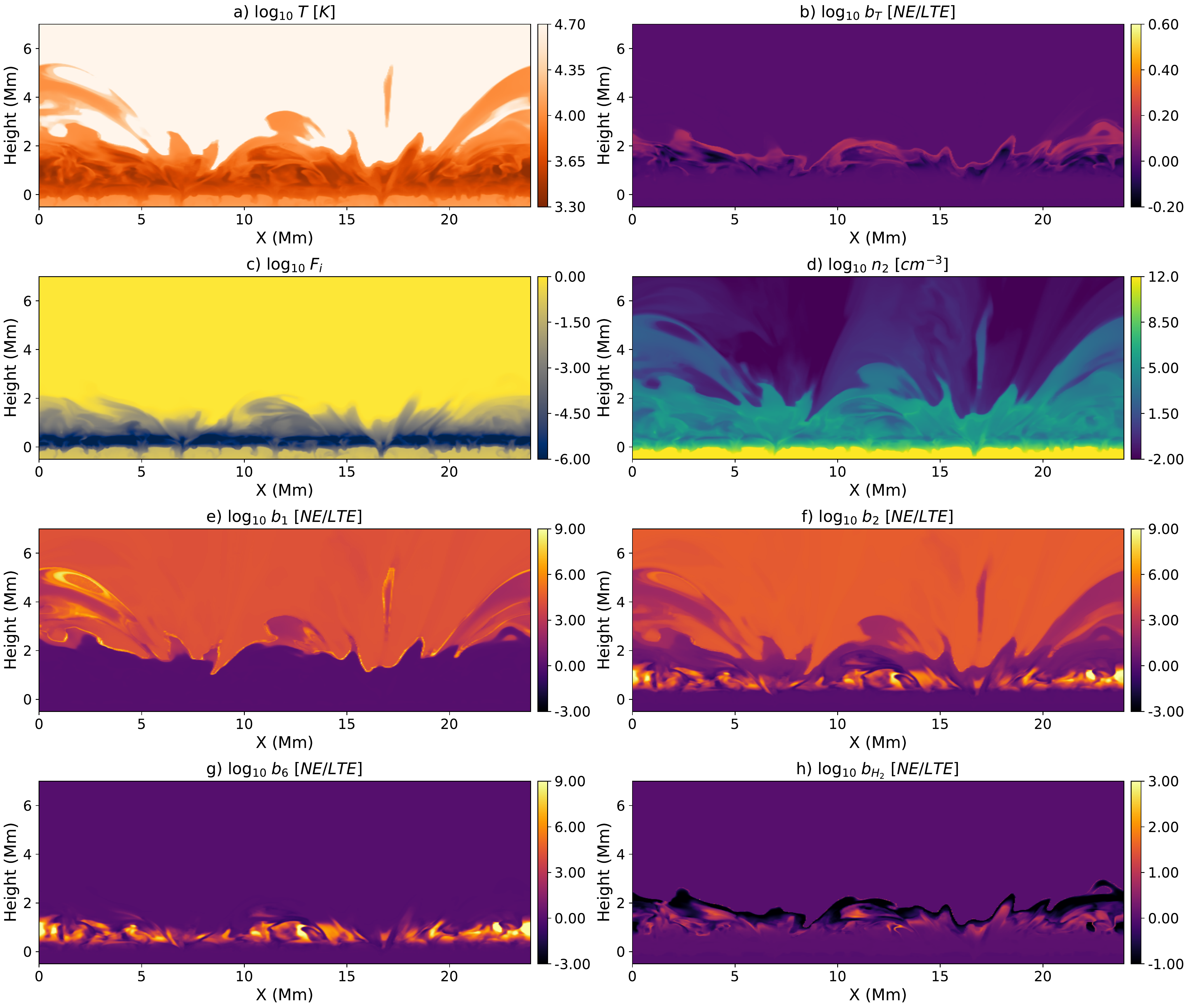}
\caption{Properties of the non-equilibrium hydrogen populations through the \textbf{centre} of the enhanced network region. The panels show a) temperature in the NE simulations, b) the departure coefficient of temperature $b_T = T_{\mathrm{NE}}/T_{\mathrm{LTE}}$, c) the NE ionisation fraction $F_i = n_{\mathrm{H},1}/n_{\mathrm{H,tot}}$, d) the number density of the first excited level of hydrogen $n_{\mathrm{H},0,1}$, and the departure coefficients of e) the ground state, f) the first excited state, g) protons and h) molecular hydrogen.}
\label{fig:Hpops_network}
\end{figure*}

A detailed look at the hydrogen populations can be seen in Fig. \ref{fig:Hpops_network}. In order to compare the NE simulation with LTE we calculate the departure coefficient $b$ of a quantity $X$ as the ratio of the value in non-equilibrium, to the value calculated using the LTE equation of state, $b_X=X_{\mathrm{NE}}/X_{\mathrm{LTE}}$.  The departure coefficient of temperature ($b_T$, panel b) shows up to $35 \%$ higher temperature in shocks and the transition region in the non-equilibrium simulation, while behind the shocks the temperature is reduced by up to $25 \%$. The ionisation fraction (panel c) is smooth throughout the chromosphere, as the long recombination timescales prevent neutral formation in the inter-shock regions. The departure coefficient for molecular hydrogen ($b_{H_2}$, panel h) is 1 in the photosphere and the temperature minimum, and around unity in cold chromospheric pockets. In warmer regions of the mid-chromosphere the departure coefficient can be temporarily enhanced, and in hot shocks and the upper chromosphere it is reduced. The increase (or decrease) in  $H_2$ departure coefficient occurs largely in locations where the temperature departure coefficient is decreased (or increased). The departure coefficients of the hydrogen ground state ($b_1$, panel e), and first excited state ($b_2$, panel f) in the corona are $b_1=1.7\times 10^4$ and $b_2=4\times 10^4$, similar to the values of \citet{leenaarts_2007_nonequilibrium}. Although small regions with $b_1>10^9$ are seen, we do not see large regions with extremely high departure coefficients just above the transition region ($b_{1}>10^{10}$), as observed in \citet{leenaarts_2007_nonequilibrium}. The departure coefficients of the Bifrost public simulation \citep{carlsson_2016_public} show similar magnitudes and behaviour as those shown in Fig. \ref{fig:Hpops_network}. They are shown for the initial snapshot, calculated from the Bifrost code, in Fig. \ref{fig:Hpops_network_IC} of Appendix \ref{sec:initial_hpops}.

\section{Numerical performance} \label{sec:performance}

\begin{table*}[htp]
    \caption{Simulation Setups}
    \centering
    \begin{tabular}{|c c c c c| c c c c c | c c c|}
    \hline
    \multicolumn{5}{|c|}{Simulation Setup} & \multicolumn{5}{c}{Timing} & \multicolumn{3}{|c|}{Computational Cost}\\
    \hline
        RT\tablefootmark{a} & EoS & $c_{\mathrm{max}}$\tablefootmark{b} & RT\tablefootmark{c} & Back & MHD\tablefootmark{d} & RT\tablefootmark{e} & Diff\tablefootmark{f} & EoS &  Total\tablefootmark{g} & dt & $\mu$s per\tablefootmark{h} & Wall time\tablefootmark{i}  \\ 
        type & type & ($\mathrm{km}\;\mathrm{s}^{-1}$) &  freq. & heating &(s) & (s) & (s) & (s) & (s) & (ms) & update & (Mcore-h)  \\
    \hline
        4-band S & NE & 2000 &  5 & on & 0.996 & 1.486 & 1.250 & 4.697 & 8.490 & 8.89 & 57.30 & 1.38 \\
        Grey A & LTE & 2000 & 5 & off & 0.645 & 0.138 & 0.048 & 0.075 & 0.980 & 8.89 & 6.61 & 0.16 \\
        Grey A & LTE & 2000 & 5 & on & 0.645 & 0.393 & 0.049 & 0.075 & 1.234 & 8.89 & 8.33 & 0.20 \\
        4-band S & LTE & 2000 & 5 &  on & 0.648 & 1.494 & 0.048 & 0.075 & 2.341 & 8.89 & 15.80 & 0.38\\
        4-band S & NE & 5000 & 5 & on & 0.997 & 1.162 & 1.565 & 4.483 & 8.270 & 3.75 & 55.81 & 3.18 \\
        4-band S & NE & 1000 & 5 & on & 0.991 & 1.619 & 1.190 & 4.616 & 8.476 & 14.0 & 57.20 & 0.87  \\
        4-band S & NE & 2000 & 1 & on & 1.024 & 4.52 & 1.206 & 4.675 & 11.50 & 8.92 & 77.61 &1.86\\
        4-band S & NE & 2000 & 10 & on & 0.995 & 0.913 & 1.210 & 4.588 & 7.786 & 8.89 & 52.55 & 1.261 \\
	    \hline
    \end{tabular}
    \label{tab:sim_setups}
    \tablefoot{
    \tablefoottext{a}{Radiation transfer (RT) calculations are either multi-band or single frequency (grey), and include scattering effects (S) or do not (A).}
    \tablefoottext{b}{The minimum value of the dynamically adjusted reduced-speed of light.}
    \tablefoottext{c}{The frequency (in iterations) at which the radiation field is updated.}
    \tablefoottext{d}{MHD routines include the calculation of the right-hand-side of the MHD equations, the div-B cleaner, and the time integration.}
    \tablefoottext{e}{RT routines include the interpolation of the opacity tables, calculation of radiation intensities and calculation of the radiative heating/cooling source term.}
    \tablefoottext{f}{Calculation of the diffusive fluxes. When the non-equilibrium module is turned on, this includes calls to the EoS in between directional sweeps.}
    \tablefoottext{g}{Other routines that contribute to the total time include boundary updates, time-step synchronisation and grid exchanges.}
    \tablefoottext{h}{Seconds taken for one core to update one grid-point.}
    \tablefoottext{i}{Million CPU hours required to calculate one solar hour.}
    }
\end{table*}

In this section we investigate the numerical cost of the newly implemented routines. These simulations were performed on the Max-Planck Computation data facilities `Raven' cluster. This cluster contains 1592 compute nodes, each consisting of Intel Xeon IceLake-SP processors (Platinum 8360Y) processors, with 72 cores run at 2.4 GHz and connected with Mellanox HDR InfiniBand network (100 Gbit/s) interconnects. For the results presented in this section we use 20 nodes, or 1440 cores. The MURaM code is written with MPI communication, and does not support hybrid shared memory calculations.

The simulation presented in this work, row 1 of table \ref{tab:sim_setups}, uses an Alfv\'en speed limit of $2000~\mathrm{km\;s^{-1}}$, has a typical time-step of $8.89~\mathrm{ms}$. We calculate the average time per iteration from 200 timestep updates, the expected computational time per grid cell update and the cost of simulating one hour of solar evolution. A summary of the timing, the time-step and the wall-time are presented in Table \ref{tab:sim_setups}.  The simulation setup shown in the paper will take $57.30~\mu \mathrm{s}$ per grid-point per core. This gives a wall time cost of approximately 1.38 million CPU-hours (Mcore-h) per hour of simulated time.

To determine the computational costs of the new physics implemented in this work we perform a number of test simulations. The simplest of these is a simulation with a LTE equation of state, grey LTE multi-group radiation transfer, and no back-heating due to coronal EUV radiation. This setup, row 2 of table \ref{tab:sim_setups}, is similar to that presented by \citet{rempel_2017_extension} utilising the coronal extension to the MURaM code and costs $0.16$ Mcore-h per hour of solar time ($6.61~\mu \mathrm{s}$ per grid-point per core). This LTE, grey simulation spends $14\%$ of the computational time on the radiation transfer (RT) modules, $66\%$ on MHD, and $5\%$ and $7.5\%$ on the EoS and diffusion treatments respectively. By comparison, in the chromospheric simulations presented in this work, the computational cost is  dominated by the EoS, and to a lesser extent the RT.

First we consider the effects of including the extended radiation transfer modules. Including the EUV back heating of the chromosphere, row 3 of table \ref{tab:sim_setups}, increases the cost of radiation transfer by almost $300\%$. This corresponds to a $26\%$ increase of the total computational time. This large increase is due to the optically thin nature of the EUV radiation in the corona, rays can cross many computational sub-domains and take more iterations to converge. Including a four-band scattering formulation, row 4 of table \ref{tab:sim_setups}, for the 3D multi-group radiation scheme further increases the computational cost of the simulation by $90\%$. Including the more realistic treatment of radiation transfer makes radiation transfer the most expensive component of the simulation, requiring $64\%$ of the computational time. This large increase is from additional iterations of the strongly scattering sub-bins. Most of the radiation groups converge quickly, in 2 or 3 iterations, similar to the grey radiation bin. However, the optically thin chromospheric lines bin can take up to 7 iterations to converge.

The greatest computational cost is the inclusion of the non-equilibrium ionisation of hydrogen in the equation of state. The new module increases the total runtime by $360\%$. This includes the solution of the Hydrogen rate equations, as well as overhead in the MHD and diffusion modules. The increase in the latter is caused by advection and diffusion of the atomic populations, and calls to the EoS to maintain consistency of the solution between directional sweeps of the diffusion routine. The simulation setup presented in this paper requires 8.6x more computational power per gridpoint than the LTE coronal simulations presented in \citet{rempel_2017_extension}.

In addition, we vary two approximations that have a significant effect on the computational cost. Firstly, we perform two simulations where the minimum limit on the speed of light is changed. Increasing the speed of light,  to $5000~\mathrm{km\;s}^{-1}$, row 5 of table \ref{tab:sim_setups},  reduces the time-step leading to a 230\% increase in the computational cost. Reducing the speed of light to $1000~\mathrm{km\;s}^{-1}$, row 6 of table \ref{tab:sim_setups},  allows a larger time-step and lowers the computational cost by 37\% .

A second choice which affects the computational cost is the frequency (in iterations) at which the radiation field is calculated. Updating the radiation field every iteration, row 7 of table \ref{tab:sim_setups}, increases the computational cost of radiation transfer by 300\%. Reducing the frequency to every 10 iterations, row 8 of table \ref{tab:sim_setups}, reduces it by 40\%. This change does not scale linearly as more iterations are needed to converge the radiation field to the required tolerance when it is calculated less regularly.

\section{Discussion and Conclusion} \label{sec:discussion_conclusion}

The current work improves on the original LTE MURaM code through the implementation of three main modules; a NE treatment of hydrogen in the EoS, NLTE tabulated losses in the chromosphere, and a scattering multi-group radiation transfer scheme. An initial simulation has been performed, beginning from the publicly available Bifrost snapshot. The simulation differs significantly from the original Bifrost model, due to differences in the diffusion scheme, the potential boundary condition, and small differences in the equation of state. A detailed comparison between the results from Bifrost and MuRAM will be the subject of a separate paper.

The simulation shows the importance of a non-equilibrium treatment of hydrogen in the chromosphere. The upper chromosphere is highly dynamic with strong shocks, and large departure coefficients of the ground state and $n=2$ energy level. Despite the strong gradients and fine structure in velocity and temperature, the hydrogen populations in the upper atmosphere are smooth due to the long recombination times, relative to the dynamical timescales. The departure coefficients $b_1$ and $b_2$ are approximately $10^3$ to $10^6$ in the upper chromosphere. The departure populations calculated match those in the Bifrost code \citep{leenaarts_2007_nonequilibrium,carlsson_2016_public}. The temperatures are around 25\% different from the LTE values and the electron number density remains higher in cold shock expansions. These differences occur due to the inability of protons to recombine to neutral hydrogen before a new shock passes through the chromosphere. These differences will be important for the accurate synthesis of chromospheric spectral lines.

The current implementation of chromospheric radiative losses and non-equilibrium equation of state are based upon a number of simplifying assumptions.
\begin{enumerate} 
\item The tabulated chromospheric line losses ignore significant scatter around pre-tabulated values of escape probability, ionisation fraction and the radiative loss function. The optically thin formalism cannot simulate 3D heating effects due to shocks and explosive events.
\item The radiation field used for the NE treatment of hydrogen is isotropic in the chromosphere.
\item The treatment of Lyman-alpha in radiative equilibrium for the NE treatment of hydrogen is inaccurate near the transition region \citep{Carlsson_2002_dynamichydrogen,golding_2016_NEhelium}.
\item Atomic and molecular populations in the multi-group radiation transport scheme are treated in LTE.
\item Helium is treated in LTE.
\end{enumerate}
The above approximations are necessary for the simulation of large 3D models including a non-equilibrium chromosphere. They will also have a significant effect on the physics and chemistry acting in the chromosphere. It is important to investigate new methods to relax these assumptions. The work by \citet{golding_2016_NEhelium} has extended the non-equilibrium EoS and chromospheric line cooling to be more accurate in the upper chromosphere and transition region. This includes an approximate Lyman-alpha bin, allowing for 3D cooling and heating effects while ignoring scattering and PRD effects. Additionally, the optically thin losses have been split into 6 EUV frequency bands to incorporate the Lyman continuum and a simplified 3-level helium atom. These improvements greatly increase the realism of the method in the upper chromospheric layers. 

Another proposal for the fast non-equilibrium treatment of atoms in a radiative MHD simulation is described by \citet{judge_2017_dynamicRT}. This method uses the escape probability approach to quickly converge the populations. The method is easily extendable to elements other than hydrogen. The 1D plane-parallel nature may lead to unrealistic variations between neighbouring horizontal pixels. It is also suggested to decouple the equation of state from atomic populations, solving the problem in stages, where the populations are updated and then used to calculate the new temperature, electron number density, and pressure. This would reduce the complexity of the system of equations that are solved and allow the use of a pre-tabulated equation of state, potentially offering a significant speed up when solving the non-equilibrium problem.

Finally, a new time-implicit numerical method for solving the detailed NE radiative MHD problem, built as a extension to the MURaM code, is presented in \citet{anusha_2021_NEMURAM}. The formulation allows larger time-steps to be taken when solving the system of rates, making the method promising for time-dependent multi-dimensional simulations.

Recent studies \citep{martinezsykora_2012_PI_1,shelyag_2016_ambiheating} have shown the importance of including ion-neutral interactions, in particular ambipolar diffusion, in simulations of the chromosphere. The collisional rates and electron number density are strongly tied to the hydrogen ionisation fraction, and this varies greatly in non-equilibrium. Including NE ionisation reduces the impact  ambipolar diffusion on the simulation, however a decreased efficiency of shock heating, and heating of cool low-lying loops are seen  \citep{martinezsykora_2020_NE_PI}. Ambipolar diffusion and the Hall effect are included in the MURaM code \citep{Cheung_2012_muram_PI,rempel_2021_efficient_nonideal} and simulations including both effects will soon be performed.

\begin{acknowledgements}

We thank the anonymous referee for suggestions that improved the paper. We would like to thank S. Danilovic for assisting with understanding of the MURaM code and facilitating collaboration with the Stockholm group. D.P. would like to acknowledge the help of I. Milic, K. Sowmya, H.N. Smitha, M. van Noort, R. Collet, and P. Judge for helpful discussion relating to NLTE and NE physics.  D.P. would also like to thank the Bifrost group for helpful explanations of the Bifrost code, especially  V. Hansteen. This project has received funding from the European Research Council (ERC) under the European Union’s Horizon 2020 research and innovation programme (grant agreement No. 695075). We gratefully acknowledge the computational resources provided by the Cobra \& Raven supercomputer systems of the Max Planck Computing and Data Facility (MPCDF) in Garching, Germany. JL was supported by a grant from the Knut and Alice Wallenberg foundation (2016.0019). This material is based upon work supported by the National Center for Atmospheric Research, which is a major facility sponsored by the National Science Foundation under Cooperative Agreement No. 1852977. LSA, VW, and AIS acknowledge support from the European Research Council (ERC) under the European Union’s Horizon 2020 research and innovation program (grant no. 715947).

\end{acknowledgements}

\bibliography{./references}
\bibliographystyle{aa}

\appendix

\section{LTE Equation of State}
\label{sec:LTE_EOS}

The ideal EoS is pre-tabulated to calculate the thermodynamic variables $n_{\mathrm{e}}$, $T$, and $p$ in terms of density $\rho$ and internal energy $\epsilon$. To start, we calculate the LTE ionisation fractions for a given density $\rho$, electron number $n_{\mathrm{e}}$, temperature $T$ and set of abundances $A_a$, using the Saha-Boltzmann equation;
\begin{equation}
\label{eqn:sahaboltzmann}
\frac{n_{a,i+1}}{n_{a,i} } = \frac{2}{n_{\mathrm{e}}}\left(\frac{2 \pi m_e k_{\mathrm{B}} T}{h^2}\right)^{1.5}\frac{U_{a,i+1}}{U_{a,i}} e^{-\chi_{a,i}/\left(k_{\mathrm{B}} T\right) + \left(Z_{a,i}+1\right) \Delta \mu},
\end{equation}
where $k_{\mathrm{B}}$ is Boltzmann's constant, $m_e$ the electron mass, h is Planck's constant and $\Delta \mu$ the ionisation lowering due to interactions with surrounding particles. $U_{a,i}$ is the partition function, $\chi_{a,i}$ the ionisation energy, $Z_{a,i}$ the charge, and and $n_{a,i}$ the number density for element $a$ and ionisation stage $i$. Where available, we use the polynomial partition functions of \citet{cardona_2005_analytic}, except for iron and nickel where the tables of \citet{halenka_2001_tables}, and \citet{halenka_2002_tables} are used. The H$^{-}$ fraction is calculated from Eqn. \ref{eqn:sahaboltzmann} and the H$_2$ and H$_2^+$ fractions are  
\begin{equation}
\label{eqn:sahaH2}
\frac{n_{H,0} n_{H,0}}{n_{\mathrm{H}_{2,0}}} = \left(\frac{2 \pi m k_{\mathrm{B}} T}{h^2}\right)^{1.5} \frac{U_{H,0}^2}{U_{\mathrm{H}_2,0}} e^{-D_{\mathrm{H}_2,0}/\left(k_{\mathrm{B}} T\right)},~\mathrm{and}
\end{equation}
\begin{equation}
\label{eqn:sahaH2p}
\frac{n_{H,0} n_{\mathrm{H},1}}{n_{H_2,1}} = \left(\frac{2 \pi m k_{\mathrm{B}} T}{h^2}\right)^{1.5} \frac{U_{H,0}}{U_{\mathrm{H}_2,1}} e^{-D_{\mathrm{H}_{2,1}}/\left(k_{\mathrm{B}} T\right)},
\end{equation}
where $m$ is the reduced mass and $D_{\mathrm{H}_2}=4.478007 \ev$ and $D_{\mathrm{H}_2^+}=2.650639 \ev$ are the dissociation energies. The partition functions of $\mathrm{H}_2$ are taken from \citep{popovas_2016_partition} and the partition functions of $\mathrm{H}_{2}^+$ are from a polynomial fit to the table of \citep{stancil_1994_continuous}. Following \citet{Mihalas_1988_MHD_eos}, the energy of the hydrogen species are
\begin{eqnarray}
E_{\mathrm{H}_2} &=& k_{\mathrm{B}} T^2 \pd{\ln U_{\mathrm{H}_2,0}}{T}, \label{eqn:E_H2}\\
E_{\mathrm{H}_2,1} &=& D_{\mathrm{H}_2,0} - D_{\mathrm{H}_2,1} + \chi_{\mathrm{H},1} + k_{\mathrm{B}} T^2 \pd{\ln U_{\mathrm{H}_2,1}}{T}, \label{eqn:E_H2p}\\
E_{\mathrm{H},-1} &=& 0.5 D_{\mathrm{H}_2} - \chi_{\mathrm{H},-1},~\mathrm{and} \label{eqn:E_Hm}\\
E_{\mathrm{H},i,e} &=& 0.5 D_{\mathrm{H}_2} + \chi_{\mathrm{H},i,e}, \label{eqn:E_H}
\end{eqnarray}
where $\chi_{\mathrm{H},i,e}$ is the excitation or ionisation energy of the hydrogen level $(i,e)$. The ionisation energy of neutral hydrogen is $\chi_{\mathrm{H},1}=13.59844\ev$, and for $\mathrm{H}^{-}$ is $\chi_{\mathrm{H},-1}=0.754\ev$. Once the fractions of all elements and of the hydrogen molecules are determined we follow the formulation of the VMW equation of state. This process involves iterating the ionisation and molecular fractions, electron number density and temperature until a convergence criteria is reached. We iterate this procedure until the electron number density converges to a tolerance of $1.0\times 10 ^{-8}$. Once converged, the energy and pressure are calculated
\begin{equation}
\label{eqn:EOSenergy}
E_{\mathrm{int}} = E_{\mathrm{exi}} + E_{\mathrm{trans}} + E_{\mathrm{rad}} + E_{\mathrm{C}} + E_{\mathrm{PI}},~\mathrm{and}
\end{equation}
\begin{equation}
\label{eqn:EOSpressure}
p = p_{\mathrm{trans}} + p_{\mathrm{rad}} + p_{\mathrm{C}} + p_{\mathrm{PI}},
\end{equation}
in terms of excitation and ionisation $(\mathrm{exi})$, translational $(\mathrm{trans})$, radiation $(\mathrm{rad})$, Coloumb $(\mathrm{C})$ and pressure ionisation $(\mathrm{PI})$ components. We ignore the effects of degenerate and relativistic electrons as they are small within the physical regime of these simulations. The radiation energy and pressure terms are also small and not included. The translational components are calculated as $p_{\mathrm{trans}} = \left(\sum_{a,i} n_{a,i} + n_{\mathrm{e}}\right)k_{\mathrm{B}} T$ and $E_{\mathrm{trans}} = 3/2\; p_{\mathrm{trans}}$.

The contribution to the energy from excitation and ionisation is calculated as
\begin{equation}
\label{eqn:internalenergy}
E_{\mathrm{exi}} = \sum_{a,i}\left(\chi_{a,i} + k_{\mathrm{B}} T^2 \pd{\ln U_{a,i}}{T}\right) n_{a,i}.
\end{equation}
For the Coulomb correction we follow the prescription of \citet{Mihalas_1988_MHD_eos}. The free energy, ignoring electron density, is calculated for a parcel of gas of volume $V$ and a particle number $N = n V$, as
\begin{equation}
\label{eqn:coulombfree}
F_{\mathrm{C}} = - \frac{2\pi^{1/2} e^3}{3 k_{\mathrm{B}}^{1/2}} \frac{1}{\left(V T\right)^{1/2}} \left(\sum_{a,i} N_{a,i} Z^2_{a,i} + N_{\mathrm{e}} \right)^{3/2} \tau\left(x\right),
\end{equation}
where the function $\tau$ is
\begin{equation}
\label{eqn:coulombtau}
\tau\left(x\right) = 3 x^{-3}\left(\ln\left(1+x\right)-x+\frac{1}{2} x^2\right),
\end{equation}
and $x$ is given by
\begin{eqnarray}
\label{eqn:coulombx}
x &=& \frac{4\pi^{1/2} e^3}{3 k_{\mathrm{B}}^{3/2}}\frac{1}{ V^{1/2} T^{3/2}}\nonumber\\
&&\frac{\left(\sum_{a,i} N_{a,i} Z_{a,i} \right)}{\left(\sum_{a,i} N_{a,i}\right)}\left(\sum_{a,i} N_{a,i} Z^2_{a,i}+N_{\mathrm{e}} \right)^{1/2}.
\end{eqnarray}
The required derivatives of $\tau$ and $x$ are
\begin{equation}
\label{eqn:dcoulombtau}
\pd{\tau}{x} = -\frac{3}{x}\left(\tau - \frac{1}{1+x}\right),
\end{equation}
\begin{equation}
\label{eqn:dcoulombxdNe}
\pd{x}{N_{\mathrm{e}}} = x \frac{1}{2 \left(\sum_{a,i} N_{a,i} Z^2_{a,i}+N_{\mathrm{e}} \right)},
\end{equation}
\begin{equation}
\label{eqn:dcoulombxdV}
\pd{x}{V} = -x \frac{1}{2 V},\mathrm{and}
\end{equation}
\begin{equation}
\label{eqn:dcoulombxdT}
\pd{x}{T} = -x \frac{3}{2 T}.
\end{equation}

The reduction of the ionisation potential $\Delta\mu_{\mathrm{C}}$, used in the Saha-Boltzmann equation (\ref{eqn:sahaboltzmann}), is 
\begin{eqnarray}
\Delta \mu_{\mathrm{C}} &=& - \frac{1}{k_{\mathrm{B}} T} \pd{F_{\mathrm{C}}}{N_{\mathrm{e}}} \label{eqn:coulombpotential}\\
&=&  -\frac{F_{\mathrm{C}}}{k_{\mathrm{B}} T} \left(\frac{3}{2}\frac{1}{\sum_{a,i} N_{a,i} Z^2_{a,i} + N_{\mathrm{e}}} + \frac{1}{\tau}\pd{\tau}{x} \pd{x}{N_{\mathrm{e}}}\right),\nonumber
\end{eqnarray}
giving a pressure correction;
\begin{equation}
\label{eqn:coulombpressure}
P_{\mathrm{C}} = - \pd{F_{\mathrm{C}}}{V} =  F_{\mathrm{C}}\left(\frac{1}{2 V} - \frac{1}{\tau}\pd{\tau}{x} \pd{x}{V}\right),
\end{equation}
and the internal energy correction;
\begin{equation}
\label{eqn:coulombenergy}
\epsilon_{\mathrm{C}} = - T^2 \pd{}{T}\left(\frac{F_{\mathrm{C}}}{T}\right) = F_{\mathrm{C}}\left(\frac{3}{2} - \frac{1}{\tau}\pd{\tau}{x}T\pd{x}{T}\right).
\end{equation}

Simplifying Eqns. \ref{eqn:coulombpressure}-\ref{eqn:coulombenergy}, and taking $V=1.0\;\mathrm{cm}^{-3}$ gives
\begin{eqnarray}
\label{eqn:coulombpotential2}
\Delta \mu_C &=&  \frac{1}{2 k_{\mathrm{B}} T}(3\tau  + \pd{\tau}{x}x)\nonumber\\
&&\frac{2\pi^{1/2} e^3}{3 k_{\mathrm{B}}^{1/2}} \frac{1}{T^{1/2}} \left(\sum_{a,i} n_{a,i} Z^2_{a,i} + n_e \right)^{1/2},
\end{eqnarray}
\begin{eqnarray}
\label{eqn:coulombpressure2}
P_C &=& - \frac{1}{2}(\tau + \pd{\tau}{x}x)\nonumber\\
&&\frac{2\pi^{1/2} e^3}{3 k_{\mathrm{B}}^{1/2}} \frac{1}{T^{1/2}} \left(\sum_{a,i} n_{a,i} Z^2_{a,i} + n_{\mathrm{e}} \right)^{3/2}, \mathrm{and}
\end{eqnarray}
\begin{eqnarray}
\label{eqn:coulombenergy2}
\epsilon_C &=& - \frac{3}{2}\left(\tau  + \pd{\tau}{x} x\right)\nonumber\\
&&\frac{2\pi^{1/2} e^3}{3 k_{\mathrm{B}}^{1/2}} \frac{1}{T^{1/2}} \left(\sum_{a,i} n_{a,i} Z^2_{a,i} + n_{\mathrm{e}} \right)^{3/2}.
\end{eqnarray}

Additionally, the pressure ionisation device described in the Eggleton, Faulkner and Flannery (EFF) EoS \citep{Eggleton_1973_EFFEoS} is included. This method provides thermodynamically consistent result that gives a qualitatively correct pressure ionisation as density increases.  
\begin{equation}
\label{eqn:PIfree}
F_{PI} = \frac{\Omega\left(T\right)}{V} \left(N_{\mathrm{e},0}^2 - N_{\mathrm{e}}^2 \right),
\end{equation}
where $N_{e,0}$ is the electron number of the gas when it is fully ionised, and $\Omega$ is
\begin{equation}
\label{eqn:PIomega}
\Omega = \frac{a_0^3}{2}\left(k_{\mathrm{B}} T + 20 \chi_0\right),
\end{equation}
where $a_0 = 5.23e^{-9}/\left<Z\right> \cm$ and $\chi_0=2.16e^{-11} \left<Z\right>^2 \erg$ and $\left<Z\right>$ is the mean charge per nucleus, and the required derivative
\begin{equation}
\label{eqn:PIdomegadT}
\pd{\Omega}{T} = \frac{a_0^3}{2}k_{\mathrm{B}}.
\end{equation}

The change in the potential can then be calculated,
\begin{equation}
\label{eqn:PIpotential}
\Delta \mu_{PI} = -\frac{1}{k_{\mathrm{B}} T} \pd{F_{PI}}{N_{\mathrm{e}}} =  \frac{2\Omega}{k_{\mathrm{B}} T} \left(T\right) n_{\mathrm{e}},
\end{equation}
the pressure correction;
\begin{equation}
\label{eqn:PIpressure}
P_{PI} = - \pd{F_{PI}}{V} =  \Omega\left(T\right) \left(n_{e,0}^2 - n_{\mathrm{e}}^2 \right),
\end{equation}
and finally the internal energy correction;
\begin{equation}
\label{eqn:PIenergy}
\epsilon_{PI} = - T^2 \pd{}{T}\left(\frac{F_{PI}}{T}\right) = \left(\Omega\left(T\right) - V T \pd{\Omega\left(T\right)}{T}\right) \left(n_{e,0}^2 - n_{\mathrm{e}}^2 \right).
\end{equation}
Once a complete solution is obtained for all required density and energy values, the entropy $s$ is calculated by integrating over the table
\begin{equation}
\label{eqn:EOSentropy}
s = \int_{\epsilon}\int_{\rho}\left(\frac{\epsilon}{T}\partial\ln\epsilon - \frac{p}{T \rho} \partial\ln\rho\right).
\end{equation}

\section{Derivatives of the Non-Equilibrium Equation of State} \label{sec:NE_EOS_ders}

EoS derivatives for chemical equilibrium of $\mathrm{H}_2^+$;
\begin{eqnarray}
\label{eqn:H2pCE_ders}
\pd{n_{\mathrm{H}_{2,1}}}{T} &=& - n_{\mathrm{H}_{2,1}}\left(\frac{3}{2T} + \frac{D_{\mathrm{H}_{2,1}}}{k_{\mathrm{B}} T^2} \right. \nonumber \\
&+& \left. \frac{1}{U_{\mathrm{H},0}} \pd{U_{H,0}}{T} - \frac{1}{U_{\mathrm{H}_{2,1}}} \pd{U_{\mathrm{H}_{2,1}}}{T} \right), \\
\pd{n_{\mathrm{H}_{2,1}}}{n_{\mathrm{H},0,j}} &=& \frac{ n_{\mathrm{H}_{2,1}}}{\sum_j n_{\mathrm{H},0,j}}, ~\mathrm{and}\\
\pd{n_{\mathrm{H}_{2,1}}}{n_{\mathrm{H},1}} &=& \frac{n_{\mathrm{H}_{2,1}}}{n_{\mathrm{H},1}},
\end{eqnarray}
And for $\mathrm{H}^{-1}$;
\begin{eqnarray}
\label{eqn:H2mCE_ders}
\pd{n_{_{\mathrm{H},-1}}}{T} &=&  - n_{_{\mathrm{H},-1}} \left( \frac{3}{2T} + \frac{\chi_{_{\mathrm{H},-1}}}{k_{\mathrm{B}} T^2} + \frac{1}{U_{H,0}} \pd{U_{\mathrm{H},0}}{T} \right), \\
\pd{n_{_{\mathrm{H},-1}}}{n_{\mathrm{e}}} &=& \frac{n_{_{\mathrm{H},-1}} }{n_{\mathrm{e}}},~\mathrm{and}  \\
\pd{n_{_{\mathrm{H},-1}}}{n_{\mathrm{H},0,j}} &=& \frac{n_{_{\mathrm{H},-1}}}{\sum_j n_{\mathrm{H},0,j}}. 
\end{eqnarray}
The derivatives for the energy conservation equation $f_0$ are
\begin{eqnarray}
\label{eqn:NE_energy_ders}
\pd{f_0}{T} &=& - \frac{1}{E_{\mathrm{int}}} \left(\frac{3 k_{\mathrm{B}}}{2}\left[n_{\mathrm{e}} + n_{\mathrm{nonH}} + n_{\mathrm{H}_2} + n_{\mathrm{H}_{2,1}} \right.\right.\nonumber\\
&+& \left. n_{_{\mathrm{H},-1}} + \sum_{i,j} n_{_{\mathrm{H},i,j}}\right] + \frac{3 k_{\mathrm{B}} T}{2}\left[\pd{n_{noH}}{T} + \pd{n_{_{\mathrm{H},-1}}}{T}\right. \nonumber\\
&+& \left. \pd{n_{_{\mathrm{H_2},1}}}{T} \right] +n_{\mathrm{H,tot}} \pd{E_{\mathrm{nonH}}}{T} + \pd{E_{H_2} }{T}n_{\mathrm{H}_2}  \nonumber\\
&+& \left.\pd{E_{_{\mathrm{H},-1}} n_{_{\mathrm{H},-1}}}{T} + \pd{E_{H_{2,1}} n_{\mathrm{H}_{2,1}}}{T}\right), \\
\pd{f_0}{n_{\mathrm{e}}} &=& - \frac{1}{E_{\mathrm{int}}} \left(\frac{3 k_{\mathrm{B}} T}{2}\left[1.0 + \pd{n_{_{\mathrm{H},-1}}}{n_{\mathrm{e}}} \right] + n_{\mathrm{H,tot}} \pd{E_{\mathrm{nonH}}}{n_{\mathrm{e}}} \right.\nonumber\\
&+&\left. \pd{n_{_{\mathrm{H},-1}}}{n_{\mathrm{e}}} E_{_{\mathrm{H},-1}}  \right), \\
\pd{f_0}{n_{\mathrm{H},0,j}} &=& - \frac{1}{E_{\mathrm{int}}} \left(\frac{3 k_{\mathrm{B}} T}{2} + E_{\mathrm{H},0,j} \right. \nonumber\\ 
&+& \left. \pd{n_{\mathrm{H}_{2,1}}}{n_{\mathrm{H},0,j}} E_{H_{2,1}} + \pd{n_{_{\mathrm{H},-1}}}{n_{\mathrm{H},0,j}} E_{_{\mathrm{H},-1}}\right),~\mathrm{and}\\
\pd{f_0}{n_{\mathrm{H},1}} &=& - \frac{1}{E_{\mathrm{int}}} \left(\frac{3 k_{\mathrm{B}} T}{2} + E_{H,1} + \pd{n_{\mathrm{H}_{2,1}}}{n_{\mathrm{H},1}} E_{H_{2,1}}   \right)
\end{eqnarray}
The derivatives for the charge conservation equation $f_1$ are
\begin{eqnarray}
\label{eqn:NE_charge_ders}
\pd{f_1}{T} &=& - \frac{1}{n_{\mathrm{e}}} \left(-\pd{n_{_{\mathrm{H},-1}}}{T} + n_{\mathrm{H,tot}} \pd{n_{\mathrm{\mathrm{e,nonH}}}}{T}\right), \\
\pd{f_1}{n_{\mathrm{e}}} &=& - \frac{1}{n_{\mathrm{e}}} \left(1 - \pd{n_{_{\mathrm{H},-1}}}{n_{\mathrm{e}}} + n_{\mathrm{H,tot}} \pd{n_{\mathrm{\mathrm{e,nonH}}}}{n_{\mathrm{e}}}\right) \\
&+& \frac{1}{n_{\mathrm{e}}^2}\left(n_{\mathrm{H},1}+n_{\mathrm{H}_{2,1}}-n_{_{\mathrm{H},-1}} + n_{\mathrm{H,tot}} n_{\mathrm{\mathrm{e,nonH}}}\right)\nonumber\\
\pd{f_1}{n_{\mathrm{H},0,j}} &=& -\frac{1}{n_{\mathrm{e}}}\left(\pd{n_{\mathrm{H}_{2,1}}}{n_{\mathrm{H},0,j}}+\pd{n_{_{\mathrm{H},-1}}}{n_{\mathrm{H},0,j}}\right),\\
\pd{f_1}{n_{\mathrm{H},1}} &=& -\frac{1}{n_{\mathrm{e}}}\left(1-\pd{n_{\mathrm{H}_{2,1}}}{n_{\mathrm{H},1}}\right).
\end{eqnarray}

The derivatives for the hydrogen nucleus conservation equation $f_2$ are
\begin{eqnarray}
\label{eqn:NE_Hcons_der}
\pd{f_2}{T} &=& - \frac{1}{n_{H,tot}}\left(1+\pd{n_{\mathrm{H}_{2,1}}}{T} + \pd{n_{_{\mathrm{H},-1}}}{T}\right), \\
\pd{f_2}{n_{\mathrm{e}}} &=& - \frac{1}{n_{H,tot}}\pd{n_{_{\mathrm{H},-1}}}{n_{\mathrm{e}}},  \\
\pd{f_2}{n_{\mathrm{H},0,j}} &=& - \frac{1}{n_{H,tot}}\left(1+\pd{n_{\mathrm{H}_{2,1}}}{n_{\mathrm{H},0,j}} + \pd{n_{_{\mathrm{H},-1}}}{n_{\mathrm{H},0,j}}\right),\\
\pd{f_2}{n_{\mathrm{H},1}} &=& - \frac{1}{n_{H,tot}}\pd{n_{\mathrm{H}_{2,1}}}{n_{\mathrm{H},1}}.
\end{eqnarray}

The derivatives of the rate-equations $f_{2+i}$ are 
\begin{eqnarray}
f_{3+ij} &=& \frac{\Delta t}{n_{aij}^{t_0}}\left(\sum_{kl \neq ij} n_{akl} \pd{P_{kl,ij}}{T} - n_{aij} \sum_{kl \neq ij} \pd{P_{ij,kl}}{T}\right.  \nonumber \\
 &+& \left. \sum_{r^+} n_{r^+,A} n_{r^+,B}n_{r^+,C} \pd{K_{r^+}}{T} \right.\nonumber\\
 &-&\left. \sum_{r^-} n_{r^-,A} n_{r^-,B}n_{r^-,C} \pd{K_{r^-}}{T} \right), \\
\pd{f_{3+ij}}{n_{\mathrm{e}}} &=& - \frac{\Delta t}{n_{aij}^o}\left(\sum_{kl \neq ij} n_{akl} \pd{P_{kl,ij}}{n_{\mathrm{e}}} \right. \nonumber\\
&-& \left. n_{aij} \sum_{kl \neq ij} \pd{P_{ij,kl}}{n_{\mathrm{e}}} \right), \\
\pd{f_{3+ij}}{n_{aij}} &=& \frac{1}{n_{aij}^o}+ \frac{\Delta t}{n_{aij}^o} \sum_{kl \neq ij} P_{ij,kl}, \\
\pd{f_{3+ij}}{n_{akl}} &=& - \frac{\Delta t}{n_{aij}} P_{kl,ij},\\
\pd{f_{3+ij}}{n_{r,A}} &=& - \frac{\Delta t}{n_{aij}} n_{r,B}n_{r,C} K_{r}.
\end{eqnarray}

\section{Rate equations for solution of non-equilibrium hydrogen} \label{sec:sollum_rates}
The radiative rates used in this work are calculated using the prescription of \cite{sollum_thesis_1999}, see also \citet{leenaarts_2006_timedependant} for a description of their implementation in a 3D simulation. The angle averaged radiation field $J_{\nu}$ for each transition is given in terms of a height-dependent radiation temperature $T_\mathrm{rad}$ by setting $J_{\nu} = B_{\nu}\left(T_{\mathrm{rad}}\right)$, where
\begin{equation}
\label{eqn:J_trans}
B_{\nu}\left(T_{\mathrm{rad}}\right) = \frac{2 h \nu_0^3}{c^2} \frac{1}{e^{h\nu/k_{\mathrm{B}} T_{\mathrm{rad}}}-1}.
\end{equation}
In the upper atmosphere $T_\mathrm{rad}$ is constant, using the values $(T_{\mathrm{Sol}})$ prescribed for each transition. These values were chosen to match comprehensive 1D RADYN simulations \citep{sollum_thesis_1999}. Below the photosphere $T_\mathrm{rad}$ is equal to the local gas temperature $T$, these are smoothly joined by setting
\begin{equation}
\label{eqn:J_trad}
J_{T_\mathrm{rad}}\left(z \right) = B_{\nu}\left(T_{\mathrm{sol}}\right) + \left[ B_{\nu}\left(T_{\mathrm{crit}}\right) - B_{\nu}\left(T_{\mathrm{sol}}\right)\right]\left(\frac{m_c\left(z\right)}{m_{\mathrm{c,crit}}}\right)^{H_{\mathrm{Sol}}}
\end{equation}
in terms of column mass $m_c$, and as well as a cutoff temperature $T_{\mathrm{crit}}$ and column mass $m_{\mathrm{c,crit}}$. The parameter $H_{\mathrm{Sol}}$ is defined for each transition, and fit to a Radyn simulation by \citet{sollum_thesis_1999}. The critical values of temperature and column mass are determined by finding the lowest point $\left(z_{\mathrm{crit}}\right)$ for which $ B_{\nu}\left(\mathrm{T}\right) = 2 B_{\nu}\left(T_\mathrm{Sol}\right) $. Below this point $T_\mathrm{rad} = T$ and above it we use Eqns. \ref{eqn:J_trans} \& \ref{eqn:J_trad} to calculate $T_{\mathrm{rad}}$. This method decouples the radiation field from the thermodynamic properties, allowing rapid calculation and fast convergence of the hydrogen populations.

The radiative rates can then be calculated from $T_{\mathrm{rad}}$ and $J_{\nu}$, see \citet{sollum_thesis_1999} for a detailed derivation. For a lower level $l$ and an upper level $u$ the upwards $R_{l,u}$ and downwards $R_{u,l}$ radiative rates are
\begin{eqnarray}
\label{eqn:sollum_boundbound}
R_{l,u} &=& \frac{4 \pi^2 e^2}{h \nu_0 m_e c} f_{l,u} J_{\nu}, \\
R_{u,l} &=& \frac{g_l}{g_u} e^{\frac{h \nu_0}{k_{\mathrm{B}} T_\mathrm{rad}}} R_{l,u}, 
\end{eqnarray}
where $e$ is the electron charge, $g_i$ the statistical weight of level i, $f_{l,u}$ is the oscillator strength, and $\nu_0$ is the line \textbf{centre} frequency. The radiation temperature $T_\mathrm{rad}=T$ when $z<z_{crit}$ and is constant when $z \geq z_{\mathrm{crit}}$. The temperature derivative of $J_{\nu}$ is 
\begin{eqnarray}
\label{eqn:J_trans_der}
\pd{J_{\nu}}{T} &=& \frac{h \nu_0}{k_{\mathrm{B}} T^2} \frac{e^{\frac{h \nu_0}{k_{\mathrm{B}} T}}}{e^{\frac{h \nu_0}{k_{\mathrm{B}} T}}-1} J_{\nu},~\mathrm{for}~ z < z_{\mathrm{crit}} \\
               &=& 0,~\mathrm{for} ~ z \geq z_{\mathrm{crit}}, \nonumber
\end{eqnarray}
and the derivatives of the bound-bound rates are 
\begin{eqnarray}
\label{eqn:sollum_boundbound_der}
\pd{R_{l,u}}{T} &=& \frac{4 \pi^2 e^2}{h \nu_0 m_e c} f_{l,u} \pd{J_{\nu}}{T},~\mathrm{for}~ z < z_{\mathrm{crit}} \\
               &=& 0,~\mathrm{for}~ z \geq z_{\mathrm{crit}},\nonumber\\
\pd{R_{u,l}}{T} &=& \frac{g_l}{g_u} \pd{R_{l,u}}{T},~\mathrm{for}~ z < z_{\mathrm{crit}} \\
               &=& 0,~\mathrm{for} ~ z \geq z_{\mathrm{crit}}. \nonumber
\end{eqnarray}

The bound-free rates between a lower level $l$ and the continuum $6$
\begin{eqnarray}
\label{eqn:sollum_boundfree}
R_{l,6} &=& \frac{8 \pi}{c^2} \alpha_0 \nu_0^3 \sum_{n=0}^{\infty} E_1\left(n \frac{h \nu_0}{k_{\mathrm{B}} T}\right), \\
R_{6,l} &=& R_{l,6} \left[\frac{n_l}{n_6}\right]_{LTE},~\mathrm{for}~ z < z_{\mathrm{crit}} \\
&=& \frac{8 \pi}{c^2} \alpha_0 \nu_0^3 \left[\frac{n_l}{n_6}\right]_{LTE} \nonumber\\
&& \sum_{n=0}^{\infty} E_1\left(\left[n \frac{T}{T_\mathrm{rad}}+  1\right] \frac{h \nu_0}{k_{\mathrm{B}} T}\right),~\mathrm{for}~ z \geq z_{\mathrm{crit}}\nonumber
\end{eqnarray}
where $\left[\frac{n_l}{n_6}\right]_{LTE}$ is the LTE population ratio, and $\alpha_0$ is the radiative absorption cross-section at the ionisation edge frequency $\nu_0$. The derivatives of the bound-free rates are
\begin{eqnarray}
\label{eqn:sollum_boundfree_der}
\pd{R_{l,6}}{T} &=& \frac{8 \pi}{c^2} \alpha_0 \nu_0^3 \frac{1}{T\left(e^{\frac{h \nu_0}{k_{\mathrm{B}} T}}\right)},~\mathrm{for}~ z < z_{\mathrm{crit}} \\
               &=& 0,~\mathrm{for}~ z \geq z_{\mathrm{crit}}\nonumber\\
\pd{R_{6,l}}{T} &=& \pd{R_{l,6}}{T}\left[\frac{n_l}{n_6}\right]_{LTE} \\
&-& R_{6,l}\left(\frac{3}{2T}+\frac{h \nu_0}{k_{\mathrm{B}} T^2}\right),~\mathrm{for}~ z < z_{\mathrm{crit}} \nonumber\\
               &=& \frac{8 \pi}{c^2} \alpha_0 \nu_0^3  \left[\frac{n_l}{n_6}\right]_{LTE} \sum_{n=0}^{\infty} \frac{\exp{\left(-\frac{ n h \nu_0}{k_{\mathrm{B}} T_\mathrm{rad}}-\frac{ h \nu_0}{k_{\mathrm{B}} T}\right)}}{n T + T_\mathrm{rad}} \frac{T}{T_\mathrm{rad}}\nonumber\\
               &-& R_{6,l}\left(\frac{3}{2T}+\frac{h \nu_0}{k_{\mathrm{B}} T^2}\right),~\mathrm{for}~ z \geq z_{crit}, \nonumber\\
\pd{R_{l,6}}{n_{\mathrm{e}}} &=& 0,\\
\pd{R_{6,l}}{n_{\mathrm{e}}} &=& \frac{R_{6,l}}{n_{\mathrm{e}}}.                
\end{eqnarray}

\begin{figure*}[htp]
\centering
\includegraphics[width=17cm]{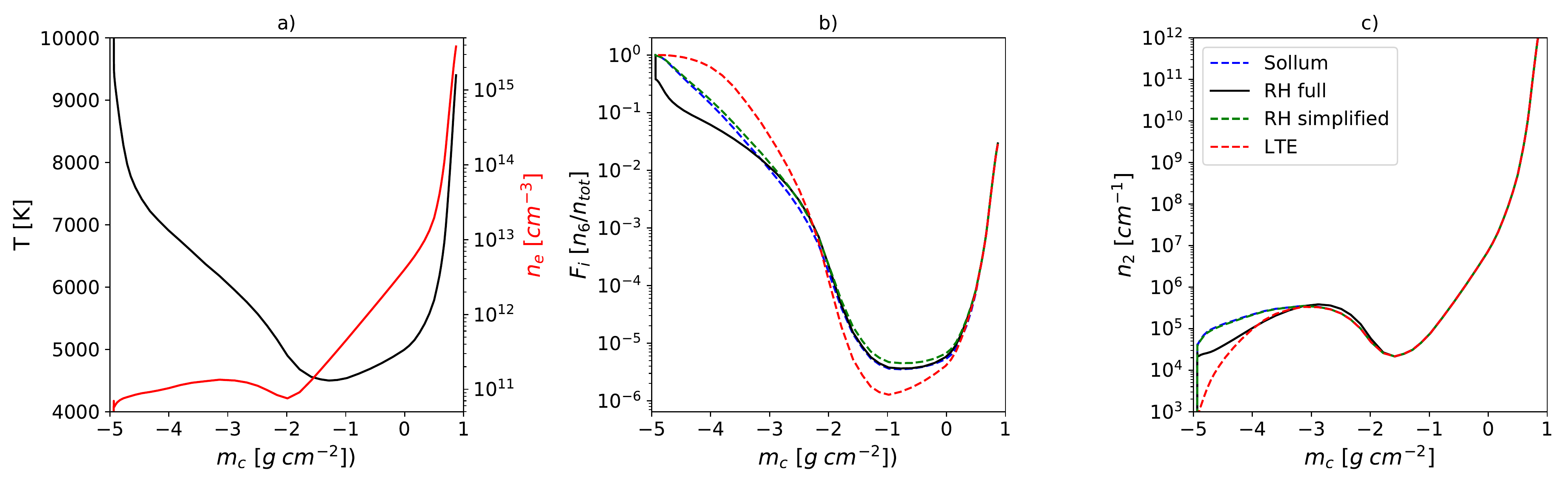}
\caption{Statistical equilibrium populations of hydrogen, comparing the Sollum radiative rates with a detailed calculation using the RH code. Panel a) shows the temperature and electron density of the FALC atmosphere. Panel b) compares the ionisation fraction, and panel c) the $n_2$ populations. These are shown for LTE (red-dash), the Sollum radiative rates (blue-dash), and from two detailed calculations with the RH code, one of which treats Lyman-lines in detailed balance (green-dash), and a second which calculates the radiative rates for the Lyman-lines (solid-black), see the legend in Panel c).}
\label{fig:RHvsSollum}
\end{figure*}

In Fig. \ref{fig:RHvsSollum} we compare detailed SE calculations made with the RH code \citep{uitenbroek_2001_RH,pereira_2015_rh1p5d} to those using the Sollum radiative rates. The statistical equilibrium solution calculated using the Sollum rates closely match the detailed solution for much of the photosphere up to the mid-chromosphere. In the upper chromosphere, the Sollum treatment of Lyman lines in detailed balance leads to a higher ionisation fraction and a lower population in the first excited state. When Lyman lines are treated in detailed balance in the RH calculation, the result closely matches that from the Sollum rates.

\section{Molecular hydrogen rates} \label{sec:Hmol_rates}
For the time dependent solution of the molecular $\mathrm{H}_2$ we use a set of rates described in table \ref{tab:Hmol_rates}, these are often described in terms of the Arrhenius equation
\begin{equation}
\label{eqn:arrhenius}
\mathrm{arr}\left(\alpha,\beta,\gamma \right) = \alpha \left(\frac{T}{300}\right)^\beta e^{-\frac{\gamma}{T}},
\end{equation}
and its derivative
\begin{equation}
\label{eqn:darrheniusdT}
\pd{\mathrm{arr}}{T}\left(\alpha,\beta,\gamma\right) = \left( \frac{\beta}{T} + \frac{\gamma}{T^2}\right) \mathrm{arr}\left(\alpha,\beta,\gamma\right).
\end{equation}
The rate coefficients, $K$, are used in the equations of molecular hydrogen and the hydrogen ground state. In the case of $K_6$ \& $K_7$ the $\mathrm{H}^{-}$ and $\mathrm{H}_2^+$ molecules are assumed to automatically dissociate.

\begin{table*}[ht]\label{tab:Hmol_rates}
\centering
\caption{\bf{Molecular hydrogen formation and dissociation rates}}

\begin{tabular}{ c | c c c }
 \hline
 $K$ & Reaction & Formula & Reference  \tabularnewline
 \hline
 $K_1$ & $H + H + H \rightarrow H_2 + H $ & $5.0e^{-32} T^{-0.25} + 2.0e^{-31} T^{-0.5}$ &  {\bf \tablefootmark{1}} \tabularnewline
 $K_2$ & $ H_2 + H + H \rightarrow H_2 + H_2$ & $\frac{1}{8} K_{HHH}$ & {\bf \tablefootmark{2}} \tabularnewline
 $K_3$ & $ H_2 + H \rightarrow H + H + H$ & $\mathrm{arr}(4.67e^{-8},-1.0,5.5e^{4})$ & {\bf UMIST 142 \tablefootmark{3}}  \tabularnewline
 $K_4$ & $ H_2 + H_2 \rightarrow H_2 + H + H$ & $\mathrm{arr}(1.0e^{-8},0,8.41e^{4})$ & {\bf UMIST 135 \tablefootmark{3}}  \tabularnewline
 $K_5$ & $H_2 + e^{-} \rightarrow H + H + e^{-}$ & $\mathrm{arr}(3.22e^{-9}, 0.35, 102000)$ & {\bf UMIST 140 \tablefootmark{3}} \tabularnewline
 $K_6$ & $H_2 + e^{-} \rightarrow H + H^{-} \rightarrow H + H + e^{-}$ & $\mathrm{arr}(1.92934e^{-11}, -1.27, 43000)$ & {\bf  \tablefootmark{4}} \tabularnewline
 $K_7$ & $H_2 + H^+ \rightarrow H_{2}^+ + H \rightarrow H + H + H^+$ & $\mathrm{arr}(1.5178e-9,-0.4563,2.1812e4)$ & {\bf \tablefootmark{5}} \tabularnewline
\hline
\end{tabular}
\tablebib{
\tablefoottext{1}{\citet{Forrey_2013_HHHrate}}
\tablefoottext{2}{\citet{palla_1983_molecules_star}}
\tablefoottext{3}{\citet{McElroy_2013_UMIST}}
\tablefoottext{4}{\citet{Hirasawa_1969_H2e_HHm}}
\tablefoottext{5}{\citet{galli_1998_chemistry}}}
\end{table*}

\section{Opacity Binning}\label{sec:scattering_opacities}

To create the group-integrated opacities for the multi-group scattering scheme we require the absorption ($\kappa$), scattering ($\sigma$) and total ($\chi=\sigma+\kappa$) opacities. The opacity contains a contribution from the continuum $^c$ and from lines $^l$. Detailed Opacity Distribution Functions (ODFs) are taken from the Merged Parallelised Simplified ATLAS code (MPS-ATLAS) \citep{MPS_ATLAS_WITZKE}, an upgraded version of ATLAS9 \citep{kurucz_1970_atlas}. For the continuum, the total and scattering opacities are available through the ATLAS package, tabulated with frequency, temperature and pressure. For the spectral line contribution, ODFs are used. The ODFs are tabulated on the same frequency, temperature and pressure grid as the continuum values, with an additional 12 sub-bins per frequency point. To extract the absorption and scattering opacity from the total opacity in the ODFs we follow \citet{skartlien_2000_multigroup}, using the approximation of \citet{vanregemorter_1962_rate}. The photon destruction probability of a frequency $\nu$ is given by the probability for collisional de-excitation from the upper level $j$ to the lower level $i$:
\begin{equation}
\epsilon^l_{\nu} \approx \frac{C_{j,i}/A_{j,i}}{1+C_{j,i}/A_{j,i}}.
\end{equation}
Here $A_{j,i}$ is the Einstein coefficient for spontaneous radiative de-excitation and $C_{j,i}$ is the collisional de-excitation parameter. Using van Regemorter's approximation
\begin{equation}
\frac{C_{j,i}}{A_{j,i}} = 20.6 \lambda^3 n_{\mathrm{e}} T^{-1/2} P\left(\frac{\Delta E_{\nu}}{k T} \right),
\end{equation}
where the function $P\left(\frac{\Delta E_{\nu}}{k T}\right)$ is pre-tabulated by \citet{vanregemorter_1962_rate}. This allows the scattering and absorption opacities to be determined using $\epsilon^l_{\nu}$
\begin{equation}
\kappa^l_{\nu} = \epsilon^l_{\nu} \chi^l_{\nu},~\mathrm{and}
\end{equation}
\begin{equation}
\sigma^l_{\nu} = (1-\epsilon^l_{\nu})\chi^l_{\nu}.
\end{equation}
The total opacities are then calculated
\begin{equation}
\kappa_{\nu} = \kappa^l_{\nu} + \kappa^c_{\nu},~\mathrm{and}
\end{equation}
\begin{equation}
\sigma_{\nu} = \sigma^l_{\nu} + \sigma^c_{\nu}.
\end{equation}

\begin{figure*}[htp]
\centering
\includegraphics[width=17cm]{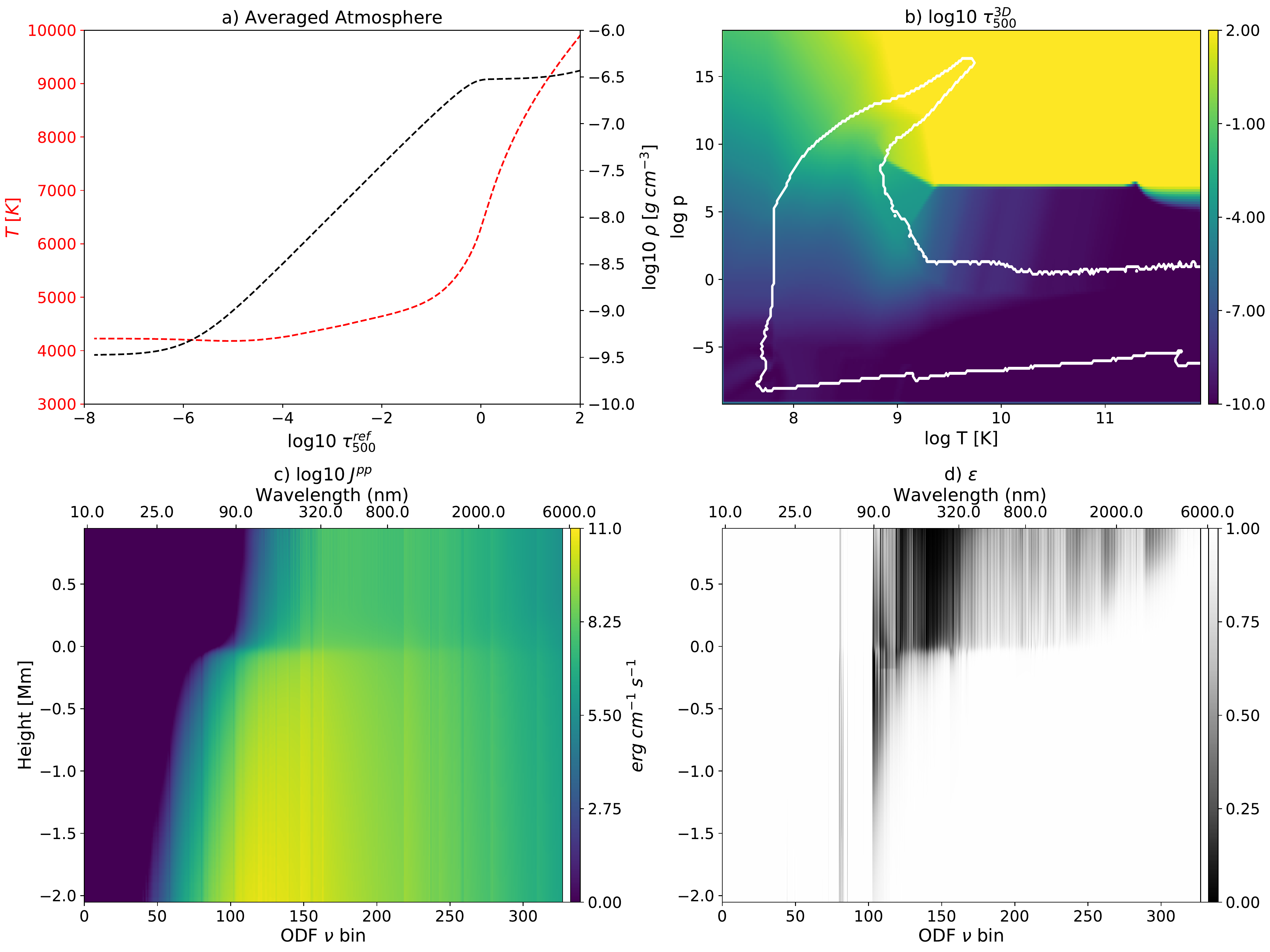}
\caption{The atmospheric properties required to calculate the multi-group scattering opacities a) The density and temperature of the reference atmosphere b) Binned optical depth $\tau_{500}^{\mathrm{3D}}$, the white contour shows the range of $T,p$ values encountered in the 3D simulation, c) Intensity $J^{\mathrm{pp}}$ used for averaging in the streaming regime and d) photon destruction probability calculated using the van-Regemorter approximation.}
\label{fig:full_ODF_RT}
\end{figure*}

 For each band j of the multigroup scheme, consisting of a set of frequencies $\Omega_j = \{\nu\}$, different averages are used to calculate the group-integrated opacities, see \citet{skartlien_2000_multigroup} for a detailed discussion. These are the Rosseland mean opacity $\chi_{j}^R$
\begin{equation}
\chi_{j}^R =\left.\int_{\Omega_j} \frac{dB_{\nu}}{dT} d\nu \middle/ \int_{\Omega_j}\frac{1}{\chi_{\nu}}\frac{dB_{\nu}}{dT} d\nu \right.,
\end{equation} 
the Planck mean opacity $\chi_{\nu}^P$
\begin{equation}
\chi_{j}^P = \left.\int_{\Omega_j} \chi_{\nu} B_{\nu}  d\nu \middle/ \int_{\Omega_j} B_{\nu} d\nu\right.,~\mathrm{and}
\end{equation}
the intensity-weighted mean $\chi_{\nu}^J$
\begin{equation}
\chi_{j}^J = \left.\int_{\Omega_j} \chi_{\nu} J_{\nu}^{\mathrm{pp}} d\nu \middle/ \int_{\Omega_j} J_{\nu}^{\mathrm{pp}} d\nu\right. .
\end{equation}
in terms of a mean intensity $J^{\mathrm{pp}}$, calculated for each bin of the ODF. This is performed using a 1D plane-parallel reference atmosphere and a short-characteristics scheme, similar to that used in the MURaM code, with $\mu=\pm 1/\sqrt{3}$ and arbitrary azimuths. In this work we use an atmosphere calculated using a column mass average of the full time-series of the Bifrost public release \citep{carlsson_2016_public}, shown in Fig. \ref{fig:full_ODF_RT}. To extrapolate from the 1D $J^{\mathrm{pp}}$ to the full range of $p,T$ a reference optical depth is used, we use $\tau_{500}$.

The extrapolation is performed following a procedure similar to that described by \citep{collet_2011_three}. For each $x,y$ pixel of each snapshot of the reference simulation time series we bin $\tau_{500}$ values for temperature and pressure corresponding to the ODF grid. Additionally we calculate a column mass averaged atmosphere of the 3D simulation. The log-mean $\tau_{500}^{\mathrm{3D}}$ is calculated for each temperature and pressure point in the table. This $\tau_{500}^{\mathrm{3D}}$ is interpolated to fill any gaps and then extrapolated to the full range of temperature and pressure used in the ODF table. The column-mass averaged background model is used to calculate $\tau^{\mathrm{ref}}_{500}$ and $J^{\mathrm{pp}}$. Using the assumption that $J^{\mathrm{pp}}$ is constant over points with the same optical depth ($\tau_{500}$) the intensity mean is calculated for the full table.
 
The resulting background model, extrapolated $\tau_{500}^{\mathrm{3D}}$, intensity $J^{\mathrm{pp}}$ and photon destruction probability $\epsilon$ can be seen in Fig. \ref{fig:full_ODF_RT}. 
 
We then follow the formalism of \citet{ludwig_thesis_1992} to join the diffusion (optically thick) and streaming (optically thin) domains. Firstly the extinction coefficient is given as
\begin{equation}
\chi_{j} = \chi_{j}^J e^{\tau_{j}/\tau_0} + \chi_{j}^R \left(1-e^{\tau_{j}/\tau_0}\right),
\end{equation}
where $\tau_0 = 0.1$ and the group optical depth $\tau_{j}$ is approximated using the expression
\begin{equation}
\tau_{j} \approx \frac{\kappa^R_{j} p}{g_{\bigodot}},
\end{equation}
where $g_{\odot}$ is the gravity at the photosphere. Similarly, the scattering albedo is
\begin{equation}
\left(1-\epsilon\right)_{j} = \frac{\sigma_{j}^J e^{\tau_{j}/\tau_0} + \sigma_{j}^P \left(1-e^{\tau_{j}/\tau_0}\right)}{\chi_{j}^J e^{\tau_{j}/\tau_0} + \chi_{j}^P \left(1-e^{\tau_{j}/\tau_0}\right)} ,
\end{equation}
and the integrated emissivity is
\begin{equation}
\left(\epsilon B\right)_{j} = \frac{\int_{\Omega_j} \kappa_{\nu} B_{\nu} d\nu}{\chi_{j}^J e^{\tau_{j}/\tau_0} + \chi_{j}^P \left(1-e^{\tau_{j}/\tau_0}\right)} .
\end{equation}

\section{Initial hydrogen populations}\label{sec:initial_hpops}

\begin{figure*}[htp]
\centering
\includegraphics[width=17cm]{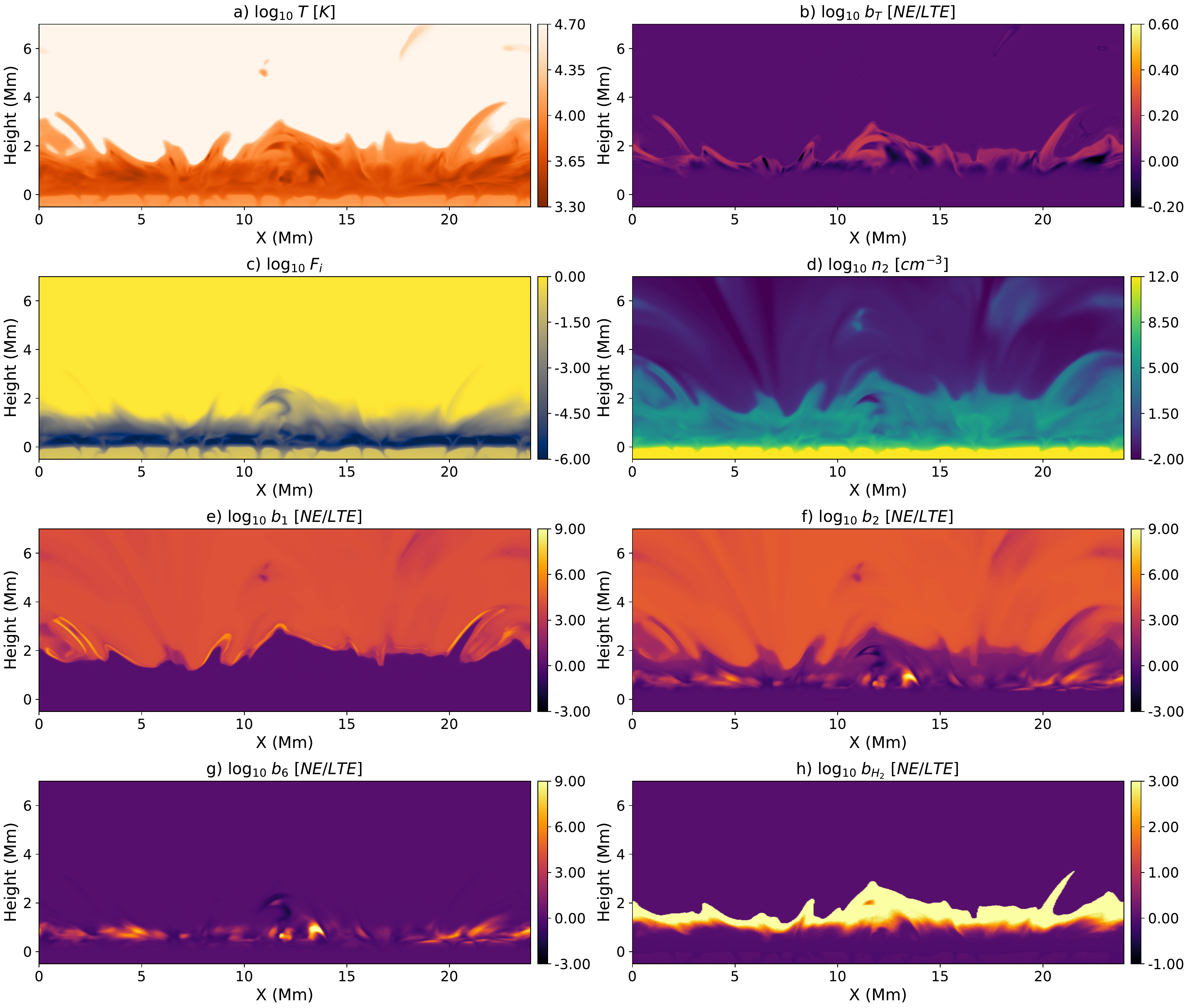}
\caption{Properties of the non-equilibrium hydrogen populations in the initial snapshot, through the \textbf{centre} of the enhanced network region. The panels show a) temperature in the NE simulations, b) the departure coefficient of temperature comparing $T_{\mathrm{LTE}}/T_{\mathrm{NE}}$, c) the NE ionisation fraction $n_{\mathrm{H},1}/n_{\mathrm{H,tot}}$, d) the first excited level of hydrogen $n_{\mathrm{H},0,1}$, and the departure coefficients of e) the ground state, f) the first excited state, g) protons and h) molecular hydrogen.}
\label{fig:Hpops_network_IC}
\end{figure*}

In Fig \ref{fig:Hpops_network} we show the departure coefficients from the initial snapshot, which was computed in Bifrost code \citep{carlsson_2016_public}. This snapshot is used as  the initial condition for the simulation presented in this paper. The departure coefficients are seen in Fig. \ref{fig:Hpops_network_IC}. The $\mathrm{H}_2$ populations are very small in the upper chromosphere, and the large departure coefficients are energetically and dynamically insignificant.

\section{Comparison to RH}\label{sec:RHSE_vs_NE}

\begin{figure*}[htp]
\centering
\includegraphics[width=17cm]{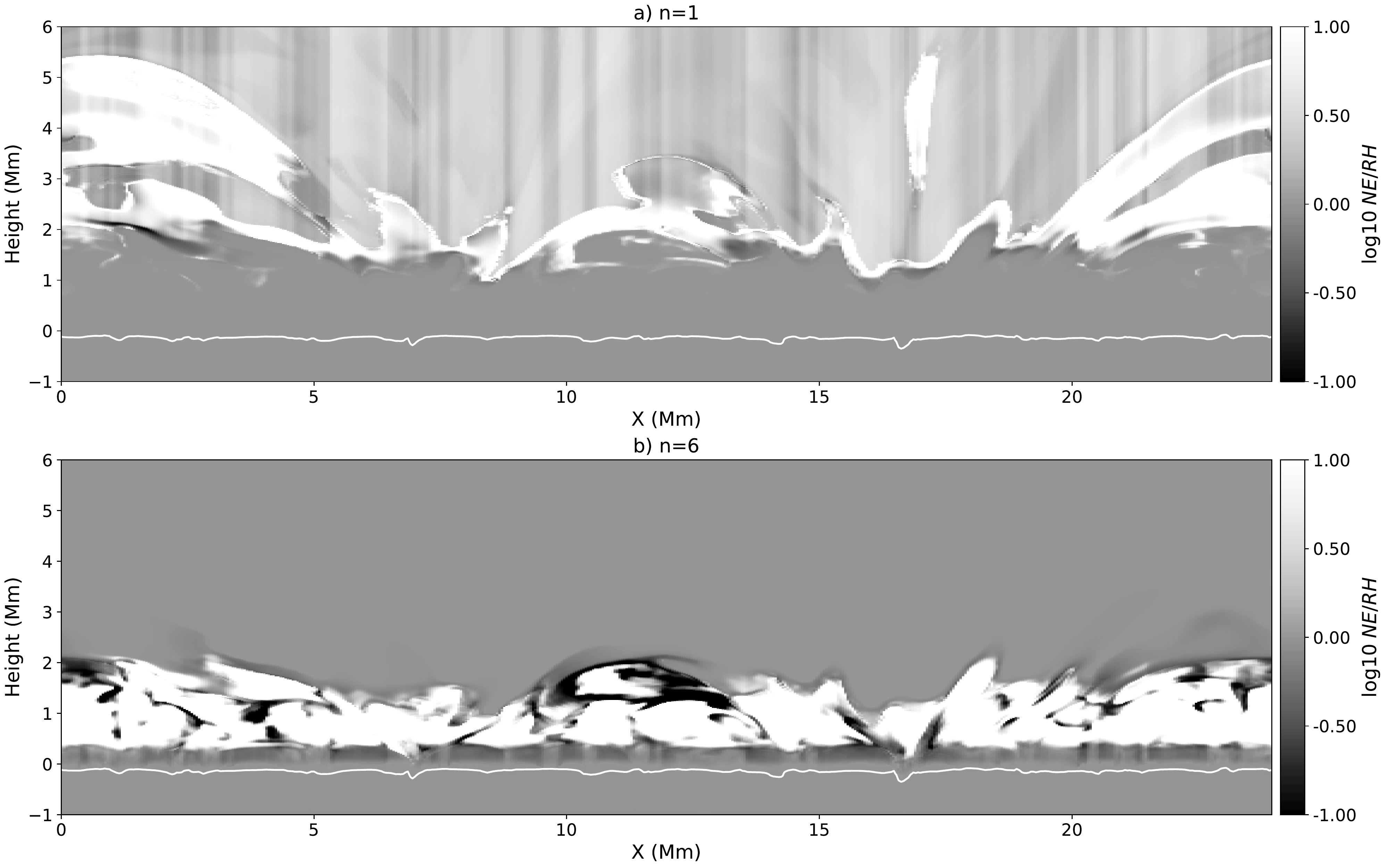}
\caption{Comparison of the ground state (panel a) and protons (panel b) from the MURaM NE treatment of hydrogen and RH.}
\label{fig:H_NEvsSE}
\end{figure*}

In order to test if the code correctly reproduces the limiting LTE and coronal equilibrium cases we reproduce the test from Sect. 11.6 of \citet{Gudiksen_2011_Bifrost}. In Fig. \ref{fig:H_NEvsSE} we compare the hydrogen populations of the MURaM NE module with those calculated in statistical equilibrium in RH \citep{uitenbroek_2001_RH,pereira_2015_rh1p5d}. The ground state population densities match the statistical equilibrium calculation for the interior and low chromosphere. In the upper-chromosphere the ground state population varies strongly from the statistical equilibrium solution, and in the corona the proton number differs, although variations remain within a factor of 10.  The proton densities match the statistical equilibrium calculation for the interior, transition region and corona. Throughout the chromosphere the proton number densities differ strongly from the statistical equilibrium solution.

\end{document}